\documentstyle[12pt]{report}
\def\thebibliography#1{\leftline{\large\it References}\list
  {[\arabic{enumi}]}{\settowidth\labelwidth{[#1]}\leftmargin\labelwidth
    \advance\leftmargin\labelsep
    \usecounter{enumi}}
    \def\newblock{\hskip .11em plus .33em minus .07em}
    \sloppy\clubpenalty4000\widowpenalty4000}

\newcommand{\be}{\begin{eqnarray}}
\newcommand{\ee}{\end{eqnarray}}
\newcommand{\dslash}{\partial \hskip -0.5em /}
\newcommand{\Dslash}{D \hskip -0.7em /}

\newcommand{\tr}{{\rm tr}}
\newcommand{\Tr}{{\rm Tr}}

\newcommand{\A}{{\cal A}}

\newcommand{\textlineskip}{\baselineskip=14pt}
\newcommand{\smalllineskip}{\baselineskip=12pt}

\begin{document}

\rightline{UNITU-THEP-16/1994}
\rightline{August 1994}
\rightline{hep-ph/9408343}
\vskip 0.8truecm
\centerline{\Large\bf Estimate of Quantum corrections to
the Mass of the}
\vskip 0.4truecm
\centerline{\Large\bf Chiral Soliton in the
Nambu--Jona--Lasinio Model}
\baselineskip=20 true pt
\vskip 0.8cm
\centerline{H.\ Weigel$^{\dag}$, R.\ Alkofer, and H.\ Reinhardt}
\vskip .3cm
\centerline{Institute for Theoretical Physics}
\centerline{T\"ubingen University}
\centerline{Auf der Morgenstelle 14}
\centerline{D-72076 T\"ubingen, Germany}
\vskip 2.0cm
\baselineskip=18pt
\centerline{\bf ABSTRACT}
\vskip 1.0cm
The Bethe--Salpeter equation for pion fluctuations off the chiral
soliton in the Nambu--Jona--Lasinio model is constructed. By
Goldstone's theorem this equation has rotational and translational
zero modes because the classical soliton is a localized stationary
field configuration which violates rotational
and translational invariance. Furthermore, the proper
normalization of the fluctuating eigen--modes is obtained.
Second quantization of the pion fluctuations off the chiral soliton
provides an energy functional of the pion fluctuations which
formally coincides with that of a harmonic oscillator.
The corresponding quantum corrections to the soliton mass together
with the semi--classical cranking prescription yield reasonable
predictions for the masses of the nucleon and the $\Delta$--resonance
when the constituent quark mass is chosen to be about 400MeV. These
calculations are, to some extend, hampered by the non--confining
character of the Nambu--Jona--Lasinio model. Comments on the $1/N_C$
counting scheme are added.

\vfill
\noindent
$^{\dag}$
{\footnotesize{Supported by a Habilitanden--scholarship of the
Deutsche Forschungsgemeinschaft (DFG).}}
\eject

\normalsize\textlineskip

\stepcounter{chapter}
\leftline{\large\it 1. Introduction}
\bigskip

In recent years it has turned out that soliton solutions of
mesonic theories provide quite a reasonable description of baryons.
These approaches are motivated by generalizing Quantum Chromo
Dynamics (QCD), which is widely believed to represent the theory
of strong interaction, to an arbitrary
number of color degrees of freedom ($N_C$) \cite{tho74}. In the
limit $N_C\rightarrow\infty$, QCD reduces to an effective theory of
weakly interating mesons (and glueballs); the effective coupling
constant being of the order $1/N_C$ \cite{tho74,wi79}. From the fact
that baryon masses increase linear in $N_C$, {\it i.e.} reciprocal
to the effective coupling constant, Witten conjectured \cite{wi79}
that baryons emerge as soliton solutions of the effective theory. It
is a common feature of solitons that their energies increase with
the inverse coupling constant. These ideas have been brought to
practice in the work of Adkins, Nappi and Witten \cite{ad83} by
reviving the Skyrme model \cite{sk61} from the early sixties. The
original Skyrme model consists of the non--linear $\sigma$--model and
an antisymmetric forth order (in derivatives) term which is mandatory
to obtain stable soliton solutions. This model has in turn been
applied to investigate many properties of baryons \cite{ho86}.
Examples are electromagnetic form factors \cite{br86} or the phase
shift analysis of pion--nucleon scattering \cite{wa84,sch89}. Over
the past decade the Skyrme model then has experienced quite an
amount of extensions\footnote{For a compilation of citations see
ref. \cite{ho93}.}. {\it E.g.} vector mesons have been added
\cite{me88} and the three flavor version of the model has been
investigated in detail\footnote{{\it Cf.} chapter III of ref.
\cite{ho93}.}. All these extensions were performed such as to comply
with the basic symmetries of QCD, in particular the chiral symmetry.
It has turned out that all these models provide a reasonable, not to
say successful, description of baryon properties. As a prominent
example one may quote the natural description of the smallness of
the matrix element of the axial singlet current, {\it i.e.} the
famous proton spin puzzle \cite{br88}. Unfortunately, one big
problem has remained unsolved over the years: The too large prediction
for the absolute mass of the nucleon. Adopting parameters which are
fixed in the meson sector as far as possible the nucleon mass is
overestimated by about 50\%, in three flavor models even more. It
has only been very recently that quantum corrections to the mass
of the Skyrmion have been evaluated \cite{mo91,ho92,ho94}. Indeed
these have been found to provide a sizable reduction of the soliton
mass leading to a predicted nucleon mass just of the right magnitude.

As a matter of fact these corrections, which are due to mesonic loops,
can be classified by the $1/N_C$--expansion. As the absolute mass of
the nucleon (938MeV) is in leading order proportional to $N_C$, the
quantum corrections are of one order less, $N_C^0$. Empirically the
size of these corrections can be estimated from the mass difference
between the nucleon and the Roper resonance (1440MeV) to be about 500MeV.
The third order in this expansion ($N_C^{-1}$) may be read off from
the difference between the nucleon mass and the position of the
$\Delta$--resonance (1232MeV). The numerical results in the Skyrme
model have been seen to follow this pattern. The mass of the Skyrmion
is found to be about 1.7--2.0GeV while quantum corrections are
somewhat less than 1GeV \cite{ho94}. The mass difference between
the nucleon and the $\Delta$--resonance is commonly employed to
adjust the only free parameter in the baryon sector of the Skyrme
model. One may summarize these studies in the Skyrme model by
stating that the problem of the too large nucleon mass in these
models has, to a large extend, been solved. What remains to be
considered are the analogue quantum corrections to other observables.

The investigations in the Skyrme model have revealed one important
fact for the evaluation of such quantum corrections \cite{ho94}: the
dominant contribution (about 90\%) is due to the zero--modes of the
soliton while the contribution of the continuum (or scattering) states
is almost negligible. These zero--modes arise because the soliton
configuration breaks the spin (isospin) and translational symmetries
of the model. As the quantum corrections are naturally subject to
subtracting the counterpart associated with the trivial vacuum, which
does not contain these zero modes but rather possesses a mass gap, it
is obvious that the zero modes may provide a sizable contribution. The
details of this issue will become more quantitative in the course of
this paper.

Only a little after the rediscovery of the Skyrme model it has been shown
that the bosonized version\cite{eb86} of the Nambu and
Jona--Lasinio (NJL) model \cite{na61,ha94} also contains soliton
solutions \cite{re88}. Although the NJL model is not as feasible as
the Skyrme model the NJL model is in a sense superior because it
represents a microscopic theory of the quark flavor dynamics. The
major purpose of the present paper is to examine the corrections to
the mass of the NJL soliton stemming from quantum fluctuations. In
many parts this will be similar to the analogous Skyrme model
calculations \cite{ho94}, however, in the NJL model new features will
arise since the meson fields here are composite fields built up from
quarks rather than being elementary as it is the case in the Skyrme
model. For example the dependence of the action on the derivative of
the meson fields with respect to the time coordinate, which is
essential for the canonical quantization, is not explicitly known. As
already mentioned it is necessary to subtract the energy of the trivial
vacuum (the baryon number zero configuration). In the NJL model this
provides a further complication because in the baryon number zero
sector the meson fields do not represent solutions of the Klein--Gordon
equation as in the Skyrme model but rather of a more involved
Bethe--Salpeter equation reflecting the quark substructure of mesons.
A further motivation for the evaluation of the quantum corrections to
the mass of the NJL soliton is the fact that its classical mass is
several hundred MeV lower than the mass of the Skyrmion. It is thus
interesting to see whether a similar reduction occurs for the
corrections and whether this will in turn lead to a good prediction for
the nucleon mass in the NJL model as well.

In order to address these questions we have organized the paper as
follows.  In the remainder of this introductory section we describe
the NJL model and the appearance of its soliton solutions. In section
2 we will introduce small amplitude fluctuations off this soliton and
briefly review the derivation of the Bethe--Salpeter equation for meson
fluctuations off the soliton. A thorough study of this equation will
provide a suitable normalization of the fluctuating modes. In contrast
to Skyrme type models this analysis will also be relevant for fluctuations
in the baryon number zero sector. In section 3 this normalization will
then prove to be useful in order to derive an energy functional for the
meson fluctuations. It will be found that these may indeed be quantized
using the formalism of second quantization. This in turn
allows us to apply techniques which have previously been developed in
the context of the Skyrme model \cite{ho94}. Section 4 is devoted to
the discussion of the meson fluctuations in the NJL model in channels
which contain zero modes. In section 5 we will numerically evaluate
the contribution of the zero modes to the quantum correction of the
soliton mass and furthermore make plausible that the contributions
of the scattering modes are negligible. Concluding remarks are left
to section 6.

As the NJL model is originally formulated in terms of quark degrees
of freedom \cite{na61} it may be understood as a microscopic model
for the quark flavor dynamics. As a matter of fact it can be
motivated in the large $N_C$ limit of QCD when the gluon propagator
is assumed to behave like a $\delta$--function in coordinate
space \cite{re90}. More importantly the NJL model exhibits the
chiral symmetry and its spontaneous breaking \cite{na61}. The latter is
reflected by a non--zero vacuum expectation value of the quark
condensate $\langle {\bar q}q\rangle$. In the process of bosonization
one introduces composite meson fields which allow one to integrate out
the quark degrees of freedom \cite{eb86}. The resulting action,
${\cal A}_{\rm NJL}$, can then be expressed as the sum
of a purely mesonic part
\be
\A_m=\int d^4x\left(-\frac{1}{4G_{\rm NJL}}
\tr(M^{\dag}M-\hat m^0(M+M^{\dag})+(\hat m^0)^2)\right)
\label{ames}
\ee
and a fermion determinant
\be
\A_F=\Tr\log(i\Dslash)=\Tr\log\left(i\dslash-(P_RM+P_LM^{\dag})\right)
\label{fdet},
\ee
{\it i.e.}
\be
\A_{NJL}=\A_F+\A_m.
\label{anjl}
\ee
Here $P_{R,L}=(1\pm \gamma _5)/2$ denote the projectors onto right--
and left--handed quark fields, respectively. As indicated in the above
expressions (\ref{ames},\ref{fdet}) we will constrains ourselves to
the investigation of scalar ($S$) and pseudoscalar ($P$) fields only.
These are parametrized by $M=S+iP$ with $S_{ij}=S^a\tau^a_{ij}/2$ and
$P_{ij}=P^a\tau^a_{ij} /2$ representing matrix fields in the two
dimensional flavor space. Alternatively one may define a polar
decomposition of the meson fields
\be
M=\xi^{\dag}_L\Sigma\xi_R.
\label{defm}
\ee
Here the matrix $\Sigma$ is Hermitian whereas the matrices
$\xi_{L,R}$ are unitary. This also gives a natural definition of the
chiral field $U=\xi^{\dag}_L\xi_R$. In eqn (\ref{ames}) $G_{\rm NJL}$
and $\hat m^0={\rm diag}\left(m^0_u,m^0_d\right)$ denote the
dimensionful NJL coupling constant and the current quark mass matrix.
For simplicity we will assume isospin symmetry, {\it i.e.}
$m^0_u=m^0_d=:m^0$.

As it stands the action is not well--defined because the fermion
determinant diverges and hence needs regularization. This is
accomplished by first continuing to Euclidean space $x_0\rightarrow
i\tau$ and then considering real and imaginary parts
\be
\A_R=\frac{1}{2}\Tr\ {\rm log}\left
(\Dslash_E\Dslash_E\hspace{0.04cm}^{\dag}\right),
\qquad
\A_I=\frac{1}{2}\Tr\ {\rm log}\left(\Dslash_E
(\Dslash_E\hspace{0.04cm}^{\dag})^{-1}\right).
\label{arai}
\ee
of the Euclidean fermion determinant separately. Here it
is important to remark that the Euclidean time, $\tau$, has to
be considered as a real quantity. Furthermore $\Dslash_E$ is the
Euclidean Dirac operator which is obtained from $\Dslash$ by the
analytic continuation described above. As a matter of fact only
${\cal A}_R$ is divergent and will be regularized by employing
Schwinger`s proper--time description \cite{sch51}. This introduces
a new parameter, the cut--off $\Lambda$, via
\be
\A_R\longrightarrow-\frac{1}{2}{\rm Tr}
\int_{1/\Lambda^2}^\infty\frac{ds}{s}\ {\rm exp}
\left(-s\Dslash_E\Dslash_E\hspace{0.04cm}^{\dag}\right).
\label{arreg}
\ee
For a sufficiently large coupling $G_{\rm NJL}$ the scalar
field possesses a non--zero vacuum expectation value
$\langle \Sigma \rangle ={\rm diag}(m,m)$ which is related to the
current quark mass $m^0$ and the quark condensate
$\langle \bar q q \rangle$ via the gap equation \cite{eb86}
\be
m & = & m^0+m^3\frac{N_C G_{\rm NJL}}{2\pi^2}
\Gamma\left(-1,(\frac{m}{\Lambda})^2\right)
=m^0-2G_{\rm NJL} \langle \bar q q \rangle.
\label{conmass}
\ee
The quantity $m$ is commonly referred to as the constituent quark mass.
Actually the non--trivial solutions to eqn (\ref{conmass}) reflect the
spontaneous breaking of the chiral symmetry by a non--vanishing
quark condensate $\langle \bar q q \rangle$.
In order to determine the parameters $G_{\rm NJL},\Lambda$ and $m$ the
fermion determinant is expanded in terms of the meson fields and
their derivatives. This then yields an effective meson theory which
allows one to express physical quantities like the pion decay constant
$f_\pi$ in terms of the cut--off $\Lambda$ and the constituent quark
mass $m$ \cite{eb86}
\be
f_\pi^2=\frac{N_C m^2}{4\pi^2}
\Gamma\left(0,\left(\frac{m}{\Lambda}\right)^2\right).
\label{fpi}
\ee
In practice the physical value $f_\pi=93$MeV is substituted in
order  to determine $\Lambda$ for a given constituent quark mass $m$.
Finally $G_{\rm NJL}$ is determined via
$G_{\rm NJL}= m^0 m/m_\pi^2f_\pi^2$ where $m_\pi=135$MeV denotes the
pion mass. Then $m$ is left as the only undetermined quantity.

In order to examine static soliton configurations it is useful to
introduce the Dirac Hamiltonian $h$ via
\be
i\beta\Dslash_E=-\partial_\tau-h.
\label{defh}
\ee
It should be noted that for static configurations only the real
part of the fermion determinant is non--zero. The functional trace
in eqn (\ref{arreg}) is performed in two consecutive steps. First
one introduces eigenstates\footnote{$\Omega_n$ are the Matsubara
frequencies. The fermionic character of the quarks requires
anti--periodic boundary conditions.} $\Omega_n=(2n+1)\pi/T$ of
$\partial_\tau$. Here $T$ denotes the Euclidean time interval under
consideration. The fermion determinant has been shown \cite{re89}
to acquire contributions from the occupation of the valence quark orbit
$-TE_{\rm val}[M]$ and the polarized Dirac sea $-TE_{\rm vac}[M]$.
Whence the total energy due to the fermion determinant is given by the
sum $E_{\rm val}[M]+E_{\rm vac}[M]-E_{\rm vac}[M=m]$,
which is a functional of the meson fields $M$. Also the subtraction
due to the trivial configuration is indicated. Defining eigenstates
of the Dirac Hamiltonian
\be
h\Psi_\nu=\epsilon_\nu\Psi_\nu
\label{diagh}
\ee
the valence quark part, $E_{\rm val}[M]$, may be expressed
as \cite{re89}
\be
E_{\rm val}[M]=N_C\sum_\nu\eta_\nu|\epsilon_\nu|
\label{eval}
\ee
wherein $\eta_\nu=0,1$ denote the valence quark occupation number
which have to be adjusted such that the total baryon number
\be
B=\sum_\nu\left(\eta_\nu-\frac{1}{2}\right)
{\rm sgn}(\epsilon_\nu).
\label{defb}
\ee
equals unity. The contribution to the energy due to the polarized
Dirac sea is obtained from ${\cal A}_R$ in the limit $T\rightarrow\infty$.
Then the temporal part of the functional trace reduces to a spectral
integral while the remainder of this trace is performed by summing
over the eigenstates of $h$. These manipulations result in \cite{re89}
\be
E_{\rm vac}[M]=\frac{N_C}{2}\int_{1/\Lambda^2}^\infty
\frac{ds}{\sqrt{4\pi s^3}}\sum_\nu{\rm exp}
\left(-s\epsilon_\nu^2\right).
\label{evac}
\ee
The classical soliton energy is finally given by the sum
\be
E_{\rm cl}[M]
=E_{\rm val}[M]+E_{\rm vac}[M]-E_{\rm vac}[M=m]+E_m[M]
\label{ecl}
\ee
wherein $E_m[M]$ originates from the mesonic part of the action
(\ref{ames}).

To be specific we employ the hedgehog {\it ansatz} for the chiral
field (in unitary gauge $\xi_L^{\dag}=\xi_R=\xi$)\footnote{Here we
attach the subscript ``$0$" for later reference when fluctuations
are included.}
\be
\xi_0(\mbox{\boldmath $r$})={\rm exp}\left(\frac{i}{2}
{\mbox{\boldmath $\tau$}} \cdot{\hat{\mbox{\boldmath $r$}}}\
\Theta(r)\right)
\label{chsol}
\ee
while the scalar fields are constrained to the chiral circle,
$\Sigma=\langle\Sigma\rangle=m$. Substituting this {\it ansatz} into
the Dirac Hamiltonian (\ref{defh}) yields
\be
h_0={\mbox {\boldmath $\alpha$}} \cdot {\mbox{\boldmath $p$}} +
\beta m \left({\rm cos}\Theta(r) + i\gamma_5{\mbox{\boldmath $\tau$}}
\cdot{\hat{\mbox{\boldmath $r$}}}\ {\rm sin}\Theta(r)\right).
\label{h0}
\ee
This Hamiltonian possesses the celebrated feature to commute with
both parity ($\hat\Pi$) and grand spin operators
($\mbox{\boldmath $G$}=\mbox{\boldmath $l$}+
\mbox{\boldmath $\sigma$}/2+\mbox{\boldmath $\tau$}/2$). The
latter is the sum of orbital angular momentum ($\mbox{\boldmath $l$}$),
spin ($\mbox{\boldmath $\sigma$}/2$) and isopin
($\mbox{\boldmath $\tau$}/2$) operators. Thus the eigenstates
$\Psi_\nu$ fall into independent subspaces which are characterized
by the eigenvalues of $\hat\Pi$ and $\mbox{\boldmath $G$}$.
For the hedgehog {\it ansatz} the mesonic part of the energy is given
by
\be
E_m[\Theta]=4\pi m_\pi^2f_\pi^2\int dr r^2
\left(1-{\rm cos}\Theta(r)\right).
\label{emtheta}
\ee

The stationary condition $\delta E_{\rm cl}[\Theta]/\delta\Theta=0$
is obtained using the chain rule. First the energy functional is
differentiated with respect to the one--particle energies
$\epsilon_\nu$. Subsequently these eigenvalues are functionally
differentiated with respect to $\Theta(r)$. This leads to the
equation of motion
\be
{\rm cos}\Theta(r)\  {\rm tr}\int d\Omega\
\rho^S(\mbox{\boldmath $r$},\mbox{\boldmath $r$})
i\gamma_5\mbox{\boldmath $\tau$} \cdot{\hat{\mbox{\boldmath $r$}}}\ =
{\rm sin}\Theta(r)\ \left\{ {\rm tr}\int d\Omega\
\rho^S(\mbox{\boldmath $r$},\mbox{\boldmath $r$})
-\frac{4\pi}{N_C}\frac{m_\pi^2 f_\pi^2}{m}\right\}
\label{eqm}
\ee
where the traces are over flavor and Dirac indices only. According to
the sum (\ref{ecl}) the scalar quark density matrix
$\rho^S(\mbox{\boldmath $x$},\mbox{\boldmath $y$}))=
\langle q(\mbox{\boldmath $x$})\bar q(\mbox{\boldmath $y$})\rangle$
is decomposed into valence quark and Dirac sea parts:
\be
\rho^S(\mbox{\boldmath $x$}, \mbox{\boldmath $y$}) & = &
\rho^S_{\rm val}(\mbox{\boldmath $x$},\mbox{\boldmath $y$})
+ \rho^S_{\rm vac}(\mbox{\boldmath $x$},\mbox{\boldmath $y$})
\nonumber \\*
\rho^S_{\rm val}(\mbox{\boldmath $x$},\mbox{\boldmath $y$}) & = &
\sum _\nu
\Psi_\nu(\mbox{\boldmath $x$})\eta_\nu
\bar \Psi_\nu(\mbox{\boldmath $y$}) {\rm sgn} (\epsilon_\nu)
\nonumber \\*
\rho^S_{\rm vac}(\mbox{\boldmath $x$},\mbox{\boldmath $y$}) & = &
\frac{-1}{2}\sum_\nu \Psi_\nu(\mbox{\boldmath $x$})
{\rm erfc}\left(\left|\frac{\epsilon_\nu}{\Lambda}\right|\right)
\bar \Psi_\nu(\mbox{\boldmath $y$}) {\rm sgn} (\epsilon_\nu) .
\label{density}
\ee
For various boundary conditions the explicit form of the
eigen--functions $\Psi_\nu(\mbox{\boldmath $x$})$ may {\it e.g.} be
found in refs. \cite{ka84,we92,al94}. Numerically the soliton is
obtained iteratively. One starts off with a test profile $\Theta(r)$
which is employed to compute the eigenvalues and --states of $h_0$
(\ref{h0}). These are then substituted into the equation of motion
(\ref{eqm}) in order to update the profile function $\Theta(r)$.
This procedure is repeated until convergence is reached.

\bigskip
\stepcounter{chapter}
\leftline{\large\it 2. Mesonic Fluctuations off the Soliton}
\bigskip

Mesonic fluctuations off the chiral soliton in the NJL model have been
examined at length in the context of the bound state approach to the
description of hyperons \cite{we93,we93a,we94}. As the general
formalism is not altered we will only point out the key steps and
emphasize the changes due to the two flavor reduction.

Here we will only consider pseudoscalar fields and constrain the scalar
fields to the chiral circle. Then a convenient parametrization for the
fluctuations off the soliton is given by \cite{we93}
\be
M(\mbox{\boldmath $r$},t)=m\
\xi_0(\mbox{\boldmath $r$})\xi_f(\mbox{\boldmath $r$},t)
\xi_f(\mbox{\boldmath $r$},t)\xi_0(\mbox{\boldmath $r$})
\label{parafl}
\ee
wherein $\xi_0(\mbox{\boldmath $r$})$ denotes static hedgehog
configuartion (\ref{chsol}) while $\xi_f(\mbox{\boldmath $r$},t)$
contains the small amplitude fluctuations
$\mbox{\boldmath $\eta$}(\mbox{\boldmath $r$},t)$
\be
\xi_f(\mbox{\boldmath $r$},t)={\rm exp}\left(\frac{i}{2}
\mbox{\boldmath $\eta$}(\mbox{\boldmath $r$},t)\cdot
\mbox{\boldmath $\tau$}\right)
=1+\frac{i}{2}\mbox{\boldmath $\eta$}(\mbox{\boldmath $r$},t)\cdot
\mbox{\boldmath $\tau$}
-\frac{1}{8}\mbox{\boldmath $\eta$}(\mbox{\boldmath $r$},t)\cdot
\mbox{\boldmath $\eta$}(\mbox{\boldmath $r$},t)+\ldots\ .
\label{xiexp}
\ee
Subsequently the total action is expanded up to quadratic order in
$\mbox{\boldmath $\eta$}(\mbox{\boldmath $r$},t)$. The zeroth order
term just renders the static energy function which is explained in the
introduction while the linear term vanishes subject to the stationary
condition (\ref{eqm}). The expression for the quadratic part has been
derived in ref.\cite{we93}. In order to present that result it is useful
to define the perturbation of the Dirac Hamiltonian
\be
i\beta\Dslash_E=-\partial_\tau-h
=-\partial_\tau-\left(h_0+h_1+h_2+\ldots\right)
\label{de2}
\ee
wherein the subscript labels the power of the meson fluctuations.
The zeroth order part is already presented in eqn (\ref{h0}).
In order to display the linear and quadratic parts we make
use of the unitary matrix \cite{eb86,re89a,we93a}
\be
{\cal T}=\xi_0 P_L + \xi_0^\dagger P_R
\label{deft}
\ee
which contains the information about the static soliton. Then
\be
h_1(\mbox{\boldmath $r$},-i\tau)=im{\cal T}\beta\gamma_5
\mbox{\boldmath $\eta$}\cdot\mbox{\boldmath $\tau$}{\cal T}^{\dag}
\qquad {\rm and}\qquad
h_2(\mbox{\boldmath $r$},-i\tau)=-\frac{m}{2}{\cal T}\beta
\mbox{\boldmath $\eta$}\cdot\mbox{\boldmath $\eta$}
{\cal T}^\dagger.
\label{h12}
\ee
Here it should be remarked that the temporal argument of the
fluctuations is continued to Euclidean space, {\it i.e.}
$\mbox{\boldmath $\eta$}=\mbox{\boldmath $\eta$}
(\mbox{\boldmath $r$},-i\tau)$
The defining equation (\ref{de2}) is then employed to expand the
operators $\Dslash_E^{\dag}\Dslash_E$ and $\Dslash_E
\left(\Dslash_E^{\dag}\right)^{-1}$ up to quadratic order in
$\mbox{\boldmath $\eta$}$. The resulting expression is subsequently
substituted into eqns (\ref{arreg}) and (\ref{arai}) for computing the
contribution of the fermion determinant to the part of the action
being quadratic in the fluctuations. At this point it is important to
note that one is forced to start off the expansion at eqn (\ref{arreg})
rather than employing standard perturbation techniques to the energy
eigenvalues (\ref{diagh}). The latter prescription would give
incorrect results since $[\partial_\tau,h_{1,2}]\ne0$. For the
expansion in Euclidean space one Fourier--transforms the fluctuations
\be
\eta_a(\mbox{\boldmath $r$},-i\tau)=\int_{-\infty}^{+\infty}
\frac{d\omega_E}{2\pi}\tilde \eta_a(\mbox{\boldmath $r$},i\omega_E)
{\rm e}^{-i\omega_E\tau}.
\label{feta}
\ee
This transformation may be transferred to the Hamiltonian:
\be
h_1(\mbox{\boldmath $r$},-i\tau)&=&\int_{-\infty}^{+\infty}
\frac{d\omega_E}{2\pi}\tilde h_{(1)}(\mbox{\boldmath $r$},i\omega_E)
{\rm e}^{-i\omega_E\tau}\quad {\rm and} \nonumber \\*
h_2(\mbox{\boldmath $r$},-i\tau) & = &
\int_{-\infty}^{+\infty}\frac{d\omega_E}{2\pi}
\int_{-\infty}^{+\infty}\frac{d\omega_E^\prime}{2\pi}
\tilde h_{(2)}(\mbox{\boldmath $r$},i\omega_E,i\omega_E^\prime)
{\rm e}^{-i(\omega_E+\omega_E^\prime)\tau}
\label{fham}
\ee
because ${\cal T}$ is static. The final expression for the harmonic
part, ${\cal A}_2$, of the Euclidean action is then continued back to
Minkowski space: $i\omega_E\rightarrow\omega$. The result may be
expressed as a sum of three pieces
\be
{\cal A}_2={\cal A}_{\rm val}+{\cal A}_{\rm vac}+{\cal A}_m.
\label{a2eta}
\ee
The contribution due to the explicit occupation of the valence
quark orbit ($|{\rm val}\rangle$) is given by
\be
{\cal A}_{\rm val}&=&-\eta^{\rm val}N_C\Bigg\{
\int^{+\infty}_{-\infty}\frac{d\omega}{2\pi}
\Big[\langle{\rm val}|
\tilde h_2(\mbox{\boldmath $r$},\omega,-\omega)|{\rm val}\rangle
\label{valquad} \\ &&\hspace{1cm}
+\sum_{\mu\ne{\rm val}}
\langle{\rm val}|\tilde h_1(\mbox{\boldmath $r$},\omega)|\mu\rangle
\langle\mu|\tilde h_1(\mbox{\boldmath $r$},-\omega)|{\rm val}\rangle
\frac{\epsilon_{\rm val}-\epsilon_\mu}
{(\epsilon_{\rm val}-\epsilon_\mu)^2-\omega^2}\Big]\Bigg\}.
\nonumber
\ee
The contribution from the polarized Dirac sea is somewhat lengthy
\be
{\cal A}_{\rm vac}& = & \frac{N_C}{2}\int_{1/\Lambda^2}^\infty
\frac{ds}{\sqrt{4\pi s}}\sum_\mu 2\epsilon_\mu
{\rm e}^{-s\epsilon_\mu^2}\int^{+\infty}_{-\infty}
\frac{d\omega}{2\pi}\langle\mu|
\tilde h_2(\mbox{\boldmath $r$},\omega,-\omega)|\mu\rangle
\nonumber \\ &&
+\frac{N_C}{4}\int_{1/\Lambda^2}^\infty ds \sqrt{\frac{s}{4\pi}}
\sum_{\mu\nu}\int^{+\infty}_{-\infty}\frac{d\omega}{2\pi}
\langle\mu|\tilde h_1(\mbox{\boldmath $r$},\omega)|\nu\rangle
\langle\nu|\tilde h_1(\mbox{\boldmath $r$},-\omega)|\mu\rangle
\nonumber
\\ && \hspace{2cm}
\times \left\{\frac{{\rm e}^{-s\epsilon_\mu^2}
+{\rm e}^{-s\epsilon_\nu^2}}{s}
+[\omega^2-(\epsilon_\mu+\epsilon_\nu)^2]
R(s;\omega,\epsilon_\mu,\epsilon_\nu) \right\}.
\label{vacquad}
\ee
The regulator function $R$ involves a Feynman parameter integral
which reflects the quark loop in the presence of the soliton
\be
R(s;\omega,\epsilon_\mu,\epsilon_\nu)
=\int_0^1 dx\ {\rm exp}\left(-s[(1-x)\epsilon_\mu^2
+x\epsilon_\nu^2-x(1-x)\omega^2]\right).
\label{regfct}
\ee
Here it should be noted that, upon explicit calculation, it
can be shown that the matrix elements $\langle\mu|
\tilde h_2(\mbox{\boldmath $r$},\omega,-\omega)|\mu\rangle$ and
$\langle\mu|\tilde h_1(\mbox{\boldmath $r$},\omega)|\nu\rangle
\langle\nu|\tilde h_1(\mbox{\boldmath $r$},-\omega)|\mu\rangle$
are invariant under $\omega\rightarrow-\omega$. This is the reason
why the terms of odd powers in $\omega$ have been dropped in the
eqns (\ref{valquad},\ref{vacquad}). Stated otherwise: The imaginary
part of the action does not contribute in the two flavor reduction
contrary to the case when strange degrees of freedom are present
\cite{we93,we93a}.  Finally the action is completed by the mesonic
part
\be
{\cal A}_m=-\frac{m_\pi^2f_\pi^2}{2}\int d^3r
\int^{+\infty}_{-\infty}\frac{d\omega}{2\pi}\
{\rm cos}\Theta\
\tilde{\mbox{\boldmath $\eta$}}(\mbox{\boldmath $r$},\omega)
\cdot\tilde{\mbox{\boldmath $\eta$}}(\mbox{\boldmath $r$},-\omega).
\label{amquad}
\ee

Formally the harmonic part of the action can be expressed with the
help of local $\Phi^{ab}_1(\mbox{\boldmath $r$})$ and bilocal
$\Phi^{ab}_2(\mbox{\boldmath $r$},\mbox{\boldmath $r$}^\prime,\omega)$
kernels
\be
{\cal A}_2&=&\frac{1}{2}\int^{+\infty}_{-\infty}\frac{d\omega}{2\pi}
\Bigg\{\int d^3r  \int d^3r^\prime
\Phi^{ab}_2(\mbox{\boldmath $r$},\mbox{\boldmath
$r$}^\prime,\omega)\tilde\eta_a(\mbox{\boldmath $r$},\omega)
\tilde\eta_b(\mbox{\boldmath $r$}^\prime,-\omega)
\nonumber \\ && \hspace{3cm}
+\int d^3r
\Phi^{ab}_1(\mbox{\boldmath $r$})
\tilde\eta_a(\mbox{\boldmath $r$},\omega)
\tilde\eta_b(\mbox{\boldmath $r$},-\omega)\Bigg\}.
\label{a2formal}
\ee
These kernels can be extracted from the above expressions. The local
kernel turns out to be diagonal in isospace
$\Phi^{ab}_1(\mbox{\boldmath $r$})=
\Phi_1(\mbox{\boldmath $r$})\delta^{ab}$. Here we recognize the
usefulness of the above defined unitary matrix ${\cal T}$ (\ref{deft})
because it considerably simplifies the presentation of these kernels
by defining chirally rotated wave--functions $\tilde \Psi_\mu=
{\cal T}^{\dag}\Psi_\mu$. Then we find
\be
\Phi_1(\mbox{\boldmath $r$})&=&-m_\pi^2f_\pi^2
{\rm cos}\Theta(r)+2\eta_{\rm val}N_Cm
\tilde \Psi^{\dag}_{\rm val}(\mbox{\boldmath $r$})\beta
\tilde \Psi_{\rm val}(\mbox{\boldmath $r$})
\nonumber \\ && \hspace{2.8cm}
-2N_Cm\int_{1/\Lambda^2}^\infty\frac{ds}{\sqrt{4\pi s}}
\sum_\mu\epsilon_\mu e^{-s\epsilon_\mu^2}
\tilde \Psi^{\dag}_\mu(\mbox{\boldmath $r$})\beta
\tilde \Psi_\mu(\mbox{\boldmath $r$})
\label{phi1}
\ee
and
\be
\Phi^{ab}_2(\mbox{\boldmath $r$},\mbox{\boldmath
$r$}^\prime,\omega)&=&
2\eta_{\rm val}N_Cm^2\sum_{\mu\ne{\rm val}}
\tilde\Psi^{\dag}_{\rm val}(\mbox{\boldmath $r$})\beta\gamma_5\tau^a
\tilde \Psi_\mu(\mbox{\boldmath $r$})
\tilde\Psi^{\dag}_\mu(\mbox{\boldmath $r$}^\prime)
\beta\gamma_5\tau^b
\tilde \Psi_{\rm val}(\mbox{\boldmath $r$}^\prime)
\frac{\epsilon_{\rm val}-\epsilon_\mu}
{(\epsilon_{\rm val}-\epsilon_\mu)^2-\omega^2}
\nonumber \\ &&
-\frac{N_C}{2}m^2\int_{1/\Lambda^2}^\infty ds\sqrt{\frac{s}{4\pi}}
\sum_{\mu\nu}
\tilde\Psi^{\dag}_\nu(\mbox{\boldmath $r$})\beta\gamma_5\tau^a
\tilde \Psi_\mu(\mbox{\boldmath $r$})
\tilde\Psi^{\dag}_\mu(\mbox{\boldmath $r$}^\prime)
\beta\gamma_5\tau^b
\tilde \Psi_\nu(\mbox{\boldmath $r$}^\prime)
\nonumber \\ && \hspace{2cm}
\times \left\{\frac{{\rm e}^{-s\epsilon_\mu^2}
+{\rm e}^{-s\epsilon_\nu^2}}{s}
+[\omega^2-(\epsilon_\mu+\epsilon_\nu)^2]
R(s;\omega,\epsilon_\mu,\epsilon_\nu) \right\}.
\label{phi2}
\ee
In ref.\cite{we93a} it has been demonstrated that these kernels are
diagonal in parity and grand spin, {\it i.e.} meson fluctuations with
different grand spin and/or parity quantum numbers decouple. This is
a consequence of the fact that the classical soliton carries zero
grand spin and has definite parity. This decoupling of the meson
fluctuations will later on be helpful since it allows us to
consider rotational and translational zero modes separately.
The equation of motion for the fluctuations is finally obtained
by varying (\ref{a2formal}) with respect to
$\tilde\eta_a(\mbox{\boldmath $r$},\omega)$
\be
\int d^3r^\prime \
\Phi^{ab}_2(\mbox{\boldmath $r$},\mbox{\boldmath
$r$}^\prime,\omega)\tilde\eta_b(\mbox{\boldmath $r$}^\prime,\omega)
+\Phi_1(\mbox{\boldmath $r$})\tilde\eta_a(\mbox{\boldmath $r$},\omega)=0
\label{bseqn}
\ee
which in fact represents the Bethe--Salpeter equation for the pion
fluctuations in the soliton background. This equation has the following
interpretation: The frequency $\omega$ has to be adjusted to $\omega_i$
such that (\ref{bseqn}) is satisfied for a non--trivial
$\tilde\eta_a(\mbox{\boldmath $r$},\omega_i)$. The frequency $\omega_i$
is called eigen--frequency and
$\tilde\eta_a(\mbox{\boldmath $r$},\omega_i)$
denotes the associated eigen--mode or --wave--function.
Below we will explain how the boundary conditions, which are imposed
on the Dirac spinors $\Psi_\nu$, transfer to the meson fluctuations
and subsequently lead to a discrete meson spectrum as well.
The numerical methods which are used to solve eqn (\ref{bseqn}) are
reported in refs.\cite{al94a,we94a}.

Later on we will have to evaluate overlap matrix elements between
the solutions to the Bethe--Salpeter eqn (\ref{bseqn}) and states
which solve this equation in the absence of the soliton {\it i.e.}
$\Theta\equiv0$. The latter states are obtained by computing the
above defined kernels $\Phi_1$ and $\Phi_2$ as mode sums which
contain the eigenvalues and --states of the free Dirac Hamiltonian
\be
h_{\rm free}=
\mbox{\boldmath $\alpha$}\cdot\mbox{\boldmath $p$}+\beta m.
\label{hfree}
\ee
Let us denote the resulting kernels by $\Phi_{1,2}^{ab(0)}$ and
the solutions to the corresponding Bethe--Salpeter equation
by $\tilde{\mbox{\boldmath $\eta$}}^{(0)}
(\mbox{\boldmath $r$},\omega_i^{(0)})$.
Of course, the eigen--frequencies $\omega_i^{(0)}$ will differ from
those obtained in the presence of the soliton. It should be remarked
that $\tilde{\mbox{\boldmath $\eta$}}^{(0)}(\mbox{\boldmath
$r$},\omega_i^{(0)})$ do not necessarily solve a Klein--Gordon equation
due to the composite nature of our meson fields.

In the first place we therefore have to find the proper normalization
of the solutions to eqn (\ref{bseqn}). In the case of the NJL model this
turns out to be more involved than {\it e.g.} for the Skyrme model
because the the frequency $\omega$ appears with all even powers in
eqn (\ref{bseqn}) rather than only quadratically. Therefore the metric
which appears in the scalar product of the underlying Hilbert space
of the fluctuations unavoidably depends on the frequency.
In order to compute this metric we (formally) expand the Bethe--Salpeter
equation (\ref{bseqn}) in terms of the frequency
\be
\sum_{n=0}^\infty\int d^3r^\prime {\cal O}^{ab}_n
(\mbox{\boldmath $r$},\mbox{\boldmath $r$}^\prime)\omega_i^{2n}
\tilde\eta_b(\mbox{\boldmath $r$}^\prime,\omega_i)=0
\label{expbs}
\ee
and assume that $\tilde\eta_a(\mbox{\boldmath $r$},\omega_i)$ represents
an eigen--mode with frequency $\omega_i$. The expansion (\ref{expbs})
obviously represents a generalization to the free Klein--Gordon
equation which corresponds to ${\cal O}_0=-m_\pi^2-\mbox{\boldmath
$\partial$}^2$ and ${\cal O}_1=1$ while all other ${\cal O}_n$ vanish.
Also in the Skyrme model only ${\cal O}_0$ and ${\cal O}_1$ are
non--zero. As compared to the Klein--Gordon equation, however, they
acquire additional space dependent factors.

For the general case we start with the identity
\be
0=\int d^3r \int d^3r^\prime
\tilde\eta_a(\mbox{\boldmath $r$},\omega_i)
\left[
{\cal O}^{ab}_0(\mbox{\boldmath $r$},\mbox{\boldmath $r$}^\prime)
-{\cal O}^{ab}_0(\mbox{\boldmath $r$},\mbox{\boldmath $r$}^\prime)
\right]\tilde\eta_b(\mbox{\boldmath $r$}^\prime,\omega_j).
\label{ortho1}
\ee
Noting that the ${\cal O}^{ab}_n$ are Hermitian under
the spatial integration we obtain by substituting the Bethe--Salpeter
equation (\ref{expbs})
\be
&&\hspace{-2cm}0=\int d^3r \int d^3r^\prime\sum_{n=1}^\infty
\tilde\eta_a(\mbox{\boldmath $r$},\omega_i)
{\cal O}^{ab}_n(\mbox{\boldmath $r$},\mbox{\boldmath $r$}^\prime)
\left[\omega_j^{2n}-\omega_i^{2n}\right]
\tilde\eta_b(\mbox{\boldmath $r$}^\prime,\omega_j)
\label{ortho2} \\
&&\hspace{-1.67cm}=\left(\omega_j^2-\omega_i^2\right)
\int d^3r \int d^3r^\prime \tilde\eta_a(\mbox{\boldmath $r$},\omega_i)
\left[\sum_{n=1}^\infty
{\cal O}^{ab}_n(\mbox{\boldmath $r$},\mbox{\boldmath $r$}^\prime)
\sum_{p=0}^{n-1}\omega_j^{2p}\omega_i^{2(n-1-p)}\right]
\tilde\eta_b(\mbox{\boldmath $r$}^\prime,\omega_j).
\label{ortho3}
\ee
Here the relation $a^n-b^n=(a-b)\sum_{m=0}^{n-1}a^mb^{(n-1-m)}$ has
been used. Eqn (\ref{ortho3}) allows one to impose the
orthonormalization condition
\be
\int d^3r \int d^3r^\prime \tilde\eta_a(\mbox{\boldmath $r$},\omega_i)
\left[\sum_{n=1}^\infty
{\cal O}^{ab}_n(\mbox{\boldmath $r$},\mbox{\boldmath $r$}^\prime)
\sum_{p=0}^{n-1}\omega_j^{2p}\omega_i^{2(n-1-p)}\right]
\tilde\eta_b(\mbox{\boldmath $r$}^\prime,\omega_j)=\delta_{ij}.
\label{ortho4}
\ee
In particular a solution $\tilde\eta_a(\mbox{\boldmath $r$},\omega_i)$ to
the Bethe--Salpeter equation is normalized to
\be
\int d^3r \int d^3r^\prime \tilde\eta_a(\mbox{\boldmath $r$},\omega_i)
{\cal M}^{ab}(\mbox{\boldmath $r$},
\mbox{\boldmath $r$}^\prime,\omega_i)
\tilde\eta_b(\mbox{\boldmath $r$}^\prime,\omega_i)=1
\label{norm1}
\ee
which defines the metric tensor
\be
{\cal M}^{ab}(\mbox{\boldmath $r$},
\mbox{\boldmath $r$}^\prime,\omega_i)
&=&\sum_{n=1}^\infty
{\cal O}^{ab}_n(\mbox{\boldmath $r$},\mbox{\boldmath $r$}^\prime)
\sum_{p=0}^{n-1}\omega_i^{2p}\omega_i^{2(n-1-p)}
=\sum_{n=1}^\infty n\ \omega_i^{2(n-1)}
{\cal O}^{ab}_n(\mbox{\boldmath $r$},\mbox{\boldmath $r$}^\prime)
\nonumber \\
&=&\frac{\partial \Phi_2^{ab}
(\mbox{\boldmath $r$},\mbox{\boldmath $r$}^\prime,\omega)}
{\partial \omega^2}\Bigg|_{\omega=\omega_i}.
\label{metric1}
\ee
As a matter of fact this metric tensor can easily be obtained
from eqn (\ref{phi2}) in contrast to the expansion coefficients
${\cal O}^{ab}_n$. We have thus succeeded in deriving a normalization
condition for the solutions to the Bethe--Salpeter equation in the
soliton background. This represents a special achievement because in
the Bethe--Salpeter equation the frequency $\omega$ appears with
arbitrary even powers. In coordinate space these are time derivatives.

As we will take advantage of the grand spin symmetry when computing
overlaps between $\tilde{\mbox{\boldmath $\eta$}}$ and
$\tilde{\mbox{\boldmath $\eta$}}^{(0)}$ it is appropriate
to also define a metric tensor
${\cal M}^{ab(0)}_i(\mbox{\boldmath $r$},
\mbox{\boldmath $r$}^\prime,\omega_i^{(0)})$
for the fluctuations in the baryon number zero sector. This quantity is
obtained by substituting $\Phi_2^{ab(0)}$ in eqn (\ref{metric1}).

Now we are finally enabled to equip the overlap
$\langle\tilde{\mbox{\boldmath $\eta$}}(\mbox{\boldmath $r$},\omega_i)|
\tilde{\mbox{\boldmath $\eta$}}^{(0)}
(\mbox{\boldmath $r$},\omega^{(0)}_j)
\rangle$ with an ingenious meaning. To do so we first redefine the
meson wave--functions $\tilde{\mbox{\boldmath $\eta$}}\rightarrow
\tilde{\mbox{\boldmath $\phi$}}$ such that they are normalized to
unity with respect to the trivial metric
\be
\int d^3r \mbox{\boldmath $\phi$}(\mbox{\boldmath $r$},\omega_i)
\cdot\mbox{\boldmath $\phi$}(\mbox{\boldmath $r$},\omega_i)=1.
\label{defphi1}
\ee
This can obviously be achieved by the introduction of the root
\cite{ho90} (see section 4 for its explicit construction)
\be
{\cal M}^{ab}(\mbox{\boldmath $r$},
\mbox{\boldmath $r$}^\prime,\omega_i)=
\int d^3 x\sum_{c=1}^3 \left(\sqrt{\cal M}\right)^{ac}
(\mbox{\boldmath $r$},\mbox{\boldmath $x$},\omega_i)\
\left(\sqrt{\cal M}\right)^{bc}
(\mbox{\boldmath $r$}^\prime,\mbox{\boldmath $x$},\omega_i)
\label{defsqrtm}
\ee
into the wave--function
\be
\phi^a(\mbox{\boldmath $r$},\omega_i)=
\int d^3 x\sum_{c=1}^3 \left(\sqrt{\cal M}\right)^{ca}
(\mbox{\boldmath $x$},\mbox{\boldmath $r$},\omega_i)\
\tilde\eta_c(\mbox{\boldmath $x$},\omega_i).
\label{defphi2}
\ee
The application of these definitions to the baryon number zero sector
is straightforward yielding
$\mbox{\boldmath $\phi$}^{(0)}(\mbox{\boldmath $r$},\omega^{(0)}_j)$.
To this end the above mentioned matrix element is defined as
\be
\langle\tilde{\mbox{\boldmath $\eta$}}
(\mbox{\boldmath $r$},\omega_i)|
\tilde{\mbox{\boldmath $\eta$}}^{(0)}
(\mbox{\boldmath $r$},\omega^{(0)}_j)\rangle:=
\int d^3r \
\mbox{\boldmath $\phi$}(\mbox{\boldmath $r$},\omega_i)\cdot
\mbox{\boldmath $\phi$}^{(0)}(\mbox{\boldmath $r$},\omega^{(0)}_j).
\label{defmatel}
\ee
This actually is the generalization of the Skyrme model definition
for this overlap \cite{ho94} in the case that the metric in the
orthogonality condition (\ref{ortho4}) depends on the frequencies
of the states. It should, however, be noted that this procedure
only yields normalized wave--functions and that there does not
exist an orthogonality condition for the fluctuations
$\mbox{\boldmath $\phi$}(\mbox{\boldmath $r$},\omega_i)$.
Nevertheless the definition (\ref{defmatel}) represents the
most reasonable one because the wave--functions
$\tilde{\mbox{\boldmath $\eta$}}$ and
$\tilde{\mbox{\boldmath $\eta$}}^{(0)}$
obey Bethe--Salpeter equations which in principle are disconnected
because they belong to different baryon sectors and thus different
Hilbert spaces.

There is one more important point which has to be mentioned in the
context of the Bethe--Salpeter equation for the meson fluctuations in
the NJL model. As this stems from a shortcoming of the NJL model in
general it effects the eigen--modes in the presence as well as in
absence of the soliton. The NJL model is well known not to possess
quark confinement. Thus meson fields may decay into ``free"
quark--antiquark pairs once the meson energy $\omega$ is beyond the
two quark threshold $\omega_{\rm thres}^{(0)}=2m$. Technically this
appears because the argument of the exponent in Feynman parameter
integral (\ref{regfct}) turns negative. As a matter of fact the
analytic continuation from Euclidean to Minkowski space yielding
eqn (\ref{vacquad}) is ill--defined for
$\omega^{(0)}\ge\omega_{\rm thres}^{(0)}$. Along this path in the
complex plane the logarithm develops an imaginary part which measures
the width for the meson to decay into a quark--antiquark pair.
Therefore the expansion of $\Phi_2^{ab(0)}$ which finally yielded
${\cal M}^{ab(0)}_i$ converges only for $\omega^{(0)}\le2m$. Later
on we will have to perform sums over the eigen--modes
$\tilde\eta^{(0)}(\mbox{\boldmath $r$},\omega_i^{(0)})$. As a consequence
of these considerations, $\omega_{\rm thres}^{(0)}$ provides a natural
cut--off to these mode sums. In the presence of the soliton the
situation is even worse because the valence quark orbit gets bound
and acquires an energy eigenvalue $\epsilon_{\rm val}<m$. Following the
analysis presented in appendix B of ref.\cite{we94} this leads to the
threshold $\omega_{\rm thres}=2|\epsilon_{\rm val}|<2m$. Fortunately,
this threshold will not be of utmost relevance for the ongoing
studies because we are mostly interested in the zero modes
($\omega_i=0$) in the soliton background.

\bigskip
\stepcounter{chapter}
\leftline{\large\it 3. Energy Functional for Fluctuations}
\bigskip

In order to compute the quantum corrections to the soliton mass we
first have to construct the energy functional for the meson
fluctuations. Again this is not straightforward because the
Bethe--Salpeter equation involves arbitrary even powers of the
frequency $\omega$. We therefore have to go back to the expansion
(\ref{expbs}) of the Bethe--Salpeter equation. Formally the
Bethe--Salpeter equation is obtained from the action formulated in
Fourier space
\be
S[\tilde{\mbox{\boldmath $\eta$}}]=\frac{1}{2}
\int \frac{d\omega}{2\pi}\int d^3r \int d^3r^\prime
\tilde\eta_a(\mbox{\boldmath $r$},-\omega)
\left[\sum_{n=0}^\infty {\cal O}^{ab}_n
(\mbox{\boldmath $r$},\mbox{\boldmath $r$}^\prime)\omega^{2n}\right]
\tilde\eta_b(\mbox{\boldmath $r$}^\prime,\omega).
\label{sfspace}
\ee
Undoing the Fourier transformation one obtains the Lagrange function
\be
L=\frac{1}{2} \int d^3r \int d^3r^\prime
\sum_{n=0}^\infty \eta_a^{(n)}(\mbox{\boldmath $r$},t)
{\cal O}^{ab}_n(\mbox{\boldmath $r$},\mbox{\boldmath $r$}^\prime)
\eta_b^{(n)}(\mbox{\boldmath $r$}^\prime,t)
\label{lagfct}
\ee
where the superscript denotes the order of the time derivative
of the fluctuation $\eta_a(\mbox{\boldmath $r$},t)$. As in
classical mechanics we compute the total derivative of $L$ with
respect to the time coordinate in order to derive the conserved
energy. Upon integration by parts we find
\be
\frac{d}{dt}L&=&\sum_{n=0}^\infty\ \int d^3r\
\Bigg\{\frac{d}{dt}\Big[\sum_{m=0}^n (-1)^m
\eta_a^{(n-m)}(\mbox{\boldmath $r$},t)
\left(\frac{\partial^m}{\partial t^m}\right)
\frac{\delta L}{\delta \eta_a^{(n)}(\mbox{\boldmath $r$},t)}\Big]
\nonumber \\
&&\hspace{3cm}
+(-1)^n\eta_a^{(n)}(\mbox{\boldmath $r$},t)
\left(\frac{\partial^n}{\partial t^n}\right)
\frac{\delta L}{\delta \eta_a^{(n)}(\mbox{\boldmath $r$},t)}\Bigg\}.
\label{dLdt}
\ee
Inserting the expansion (\ref{lagfct}) and the
Fourier--transformation for the fluctuations (\ref{feta}) the
last, {\it i.e.} surface term in (\ref{dLdt}) is shown to vanish
for meson fields which satisfy the Bethe--Salpeter equation
(\ref{bseqn},\ref{expbs}). Thus the conserved quantity which has to be
identified as the energy functional and thus the Hamiltonian is given by
\be
{\cal H}[\mbox{\boldmath $\eta$}]&=&
\int d^3r\ \left\{\sum_{n=0}^\infty \sum_{m=0}^n (-1)^m
\eta_a^{(n-m)}(\mbox{\boldmath $r$},t)
\left(\frac{\partial^m}{\partial t^m}\right)
\frac{\delta L}{\delta \eta_a^{(n)}(\mbox{\boldmath $r$},t)}\right\}-L
\nonumber \\
&=&\int d^3r \int d^3r^\prime \ \Bigg\{
\sum_{n=0}^\infty \sum_{m=0}^n (-1)^m
\eta_a^{(n-m)}(\mbox{\boldmath $r$},t)
{\cal O}^{ab}_n(\mbox{\boldmath $r$},\mbox{\boldmath $r$}^\prime)
\eta_b^{(n+m)}(\mbox{\boldmath $r$}^\prime,t)
\nonumber \\ &&\hspace{3cm}
-\frac{1}{2}\sum_{m=0}^n
\eta_a^{(n)}(\mbox{\boldmath $r$},t)
{\cal O}^{ab}_n(\mbox{\boldmath $r$},\mbox{\boldmath $r$}^\prime)
\eta_b^{(n)}(\mbox{\boldmath $r$}^\prime,t)\Bigg\}.
\label{efunct}
\ee
Next the fluctuating field
$\mbox{\boldmath $\eta$}(\mbox{\boldmath $r$},t)$
is decomposed in terms of the solutions to the
Bethe--Salpeter equation (\ref{bseqn})
\be
\mbox{\boldmath $\eta$}(\mbox{\boldmath $r$},t)=\sum_i
\frac{1}{\sqrt{2\omega_i}}\left\{
a_i\tilde{\mbox{\boldmath $\eta$}}(\mbox{\boldmath $r$},\omega_i)
e^{i\omega_i t}+
a_i^{\dag}\tilde{\mbox{\boldmath $\eta$}}(\mbox{\boldmath $r$},\omega_i)
e^{-i\omega_i t}\right\}
\label{etadecomp}
\ee
because the solutions to (\ref{bseqn}) come in pairs $\pm\omega_i$,
{\it i.e.} $\mbox{\boldmath $\eta$}(\mbox{\boldmath $r$},t)$ describes
a real field.  The canonical quantization prescription then corresponds
to require the commutation relations
\be
\left[a_i,a_j^{\dag}\right]=\delta_{ij}.
\label{comrel}
\ee
The decomposition (\ref{etadecomp}) is substituted into the
energy functional for the meson fluctuations (\ref{efunct}).
The aim is then to obtain an expression in terms of the operators
$a_i$ and $a_i^{\dag}$ by using the orthonormalization condition
(\ref{ortho4}). It is obvious that at an intermediate step
the term involving ${\cal O}^{ab}_0$ has to be eliminated. This
is done in a symmetric way by the help of the Bethe--Salpeter
equation
\be
&&\int d^3r \int d^3r^\prime \
\tilde\eta_a(\mbox{\boldmath $r$},\omega_i)
{\cal O}^{ab}_0(\mbox{\boldmath $r$},\mbox{\boldmath $r$}^\prime)
\tilde\eta_b(\mbox{\boldmath $r$}^\prime,\omega_j)=
\nonumber \\ && \hspace{2cm}
-\frac{1}{2} \int d^3r \int d^3r^\prime \
\tilde\eta_a(\mbox{\boldmath $r$},\omega_i)
\left[\sum_{n=1}^\infty\left(\omega_i^{2n}+\omega_j^{2n}\right)
{\cal O}^{ab}_n(\mbox{\boldmath $r$},\mbox{\boldmath $r$}^\prime)
\tilde\eta_b(\mbox{\boldmath $r$}^\prime,\omega_j)\right].
\label{elo0}
\ee
After a somewhat tedious calculation we find
\be
{\cal H}&=&\frac{1}{2}\sum_i\omega_i\int d^3r \int d^3r^\prime \
\tilde\eta_a(\mbox{\boldmath $r$},\omega_i)
\left[\sum_{n=1}^\infty n\ \omega_i^{2(n-1)}
{\cal O}^{ab}_n(\mbox{\boldmath $r$},\mbox{\boldmath $r$}^\prime)\right]
\tilde\eta_b(\mbox{\boldmath $r$}^\prime,\omega_i)
\nonumber \\
&&+\sum_{ij}\frac{1}{\sqrt{4\omega_i\omega_j}}
\int d^3r \int d^3r^\prime \
\tilde\eta_a(\mbox{\boldmath $r$},\omega_i)
\left[\sum_{n=1}^\infty\sum_{m=0}^{n-1}
\omega_i^{2m}\omega_j^{2(n-m-1)}
{\cal O}^{ab}_n(\mbox{\boldmath $r$},\mbox{\boldmath $r$}^\prime)\right]
\tilde\eta_b(\mbox{\boldmath $r$}^\prime,\omega_j)
\nonumber \\ && \hspace{2cm}
\times\Bigg[\frac{a_ia_j}{4}
\left(\omega_i-\omega_j\right)^2e^{i(\omega_i+\omega_j)t}
+\frac{a_i^{\dag}a_j^{\dag}}{4}
\left(\omega_i-\omega_j\right)^2e^{-i(\omega_i+\omega_j)t}
\nonumber \\ && \hspace{8cm}
+\frac{a_i^{\dag}a_j}{2}
\left(\omega_i+\omega_j\right)^2e^{-i(\omega_i-\omega_j)t}\Bigg].
\label{eintermed}
\ee
Here we finally recognize the appearance of the orthonormalization
condition (\ref{ortho4}). Hence we arrive at the energy operator of an
harmonic oscillator
\be
{\cal H}=\sum_i\omega_i\left(a_i^{\dag}a_i+\frac{1}{2}\right).
\label{eharmosc}
\ee
Although this result is not unexpected it is at the same time
non--trivial because the Bethe--Salpeter equation (\ref{bseqn}),
which is the defining equation for the normal modes of energy
$\omega_i$, contains arbitrary even powers of the time derivative
operator when transformed to coordinate space. Needless to mention
that the above analysis goes through as well in the absence of the
soliton resulting in
\be
{\cal H}^{(0)}=\sum_i\omega^{(0)}_i
\left(a_i^{(0)\dag}a^{(0)}_i+\frac{1}{2}\right).
\label{eharmosc0}
\ee
Here $a_i^{(0)\dag}$ and $a^{(0)}_i$ respectively denote the
creation and annihilation operators for the eigen--modes
$\tilde{\mbox{\boldmath $\eta$}}^{(0)}
(\mbox{\boldmath $r$},\omega^{(0)}_i)$.

We have thus seen that the vacuum energy, $\sum_i \omega_i/2$,
corresponds to the one of an harmonic oscillator and that this
result is independent of the specific form of the kernel for the
Bethe--Salpeter equation. There are two restrictions only:

\vskip0.5cm
(i) The eigenvalues appear in pairs $\pm\omega_i$.\hfil\break

(ii) The background field is static.\hfil\break
\vskip0.5cm
\noindent
In order to see that the vacuum contribution to the
Hamiltonian (\ref{eharmosc}) comes with no surprise but rather is
a feature of the semi--classical treatment of the meson fluctuations
we read off the inverse propagator for these fluctuations from the
action (\ref{sfspace})
\be
\left({\cal D}^{-1}\right)^{ab}(\mbox{\boldmath $r$},
\mbox{\boldmath $r$}^\prime,\omega^2)=\sum_{n=0}^\infty
{\cal O}^{ab}_n (\mbox{\boldmath $r$},
\mbox{\boldmath $r$}^\prime)\omega^{2n}.
\label{defprop}
\ee
Since the background field is static the propagator ${\cal D}$ is
local rather than bilocal in the frequency $\omega$.

In order to extract the vacuum contribution of the meson fluctuations
in the soliton background in the semi--classical approximation one
considers the functional integral (continued to Euclidean space)
\be
e^{-{\cal A}_M}=\int D\tilde\eta\
{\rm exp}\left\{-S_E[\tilde \eta]\right\}
=\left[{\rm Det}\left({\cal D}^{-1}\right)\right]^{-\frac{1}{2}},
\label{bosdet1}
\ee
{\it i.e.}
\be
{\cal A}_M=\frac{1}{2}\ {\rm Tr}\
{\rm log}\left({\cal D}^{-1}\right).
\label{bosdet2}
\ee
The temporal part of the functional trace can be done because
the propagator is local in the frequency. As in the case for the
fermion determinant ({\it cf}. section 2) the vacuum part is
obtained by considering the limit of infinitely large Euclidean
times $T$. Then the temporal part of the trace becomes a spectral
integral
\be
{\cal A}_M=\frac{T}{2} \int_{-\infty}^\infty \frac{d\omega}{2\pi}\
{\it Tr}\ {\rm log}\left({\cal D}^{-1}(\omega^2)\right)
\label{bosdet3}
\ee
where we have suppressed spatial and isospin arguments. Furthermore
${\it Tr}$ refers to the trace over all degrees of freedom other
than the time coordinate. Next we integrate by parts
\be
{\cal A}_M=-\frac{T}{2}\int_{-\infty}^\infty\frac{d\omega}{2\pi}\
{\it Tr}\ \left(2\omega^2 {\cal D}(\omega^2)
\frac{\partial}{\partial\omega^2}{\cal D}^{-1}(\omega^2)\right).
\label{bosdet4}
\ee
The surface term has disappeared since the propagator only depends
on $\omega^2$. The trace ${\it Tr}$ can be evaluated with the help
of the solutions to the Bethe--Salpeter equation
$\tilde{\mbox{\boldmath $\eta$}}(\mbox{\boldmath $r$},\omega_i)$
\be
{\cal A}_M=\frac{T}{2} \int_{-\infty}^\infty
\frac{d\omega}{2\pi}\ 2\omega^2 \sum_i
\langle i|{\cal D}(\omega^2)\frac{\partial}{\partial\omega^2}
{\cal D}^{-1}(\omega^2)|i\rangle.
\label{bosdet5}
\ee
Now it is important to note that the states $|i\rangle$ represent
the solutions to the Bethe--Salpeter equation. This allows one
to expand
\be
\langle i|{\cal D}(\omega^2)=\frac{1}{\omega^2-\omega_i^2}
\left\{\frac{\partial}{\partial\omega^2}
\langle i|{\cal D}^{-1}(\omega^2)\Big|_{\omega=\omega_i}
+{\cal O}\left(\omega^2-\omega_i^2\right)\right\}^{-1}.
\label{bosdet6}
\ee
Hence
\be
{\cal A}_M&=&-\frac{T}{2} \int \frac{d\omega}{2\pi}\
\sum_i\frac{2\omega^2}{\omega^2-\omega_i^2}
\nonumber \\ && \hspace{1.5cm}\times
\langle i|\left\{\frac{\partial}{\partial\omega^2}
{\cal D}^{-1}(\omega^2)\Big|_{\omega=\omega_i}
+{\cal O}\left(\omega^2-\omega_i^2\right)\right\}^{-1}
\frac{\partial}{\partial\omega^2}
{\cal D}^{-1}(\omega^2)|i\rangle.
\label{bosdet7}
\ee
Under the assumption that summation and integration may be exchanged
the spectral integral may be computed with the help of the
residue theorem
\be
{\cal A}_M&=&-\frac{T}{2}\sum_i\lim_{\omega\to\omega_i}
\left(\omega-\omega_i\right)
\frac{2\omega^2}{\omega^2-\omega_i^2}
\langle i|\left\{\frac{\partial}{\partial\omega^2}
{\cal D}^{-1}(\omega^2)\Big|_{\omega=\omega_i}\right\}^{-1}
\frac{\partial}{\partial\omega^2}
{\cal D}^{-1}(\omega^2)|i\rangle
\nonumber \\
&=&-\frac{T}{2}\sum_i\omega_i.
\label{bosdet8}
\ee
Thus we have seen on the formal level that the above listed
restrictions to the Bethe--Salpeter equation are sufficient to
provide a vacuum energy which is given by a sum of eigenfrequencies.
 However, the above derivation has to be taken
with some care because the functional traces are, of course,
divergent and thus require regularization as do the expressions
(\ref{eharmosc}) and (\ref{eharmosc0}). This issue will be discussed
next.

The main content of the above considerations is indeed the
fact that we have obtained a quantized Hamiltonian for the
meson fluctuations which are formally equivalent to a harmonic
oscillator. Considering eqn (\ref{bosdet8}) one might be tempted to
identify the quantum corrections to the energy  as the difference
\be
\frac{1}{2}\sum_i\omega_i-\frac{1}{2}\sum_j\omega_j^{(0)}.
\nonumber
\ee
However, this expression diverges logarithmically. In the context
of the Skyrme model Holzwarth has shown that two additional
subtractions yield a finite (renormalized) result \cite{ho94}.
To identify these subtractions in the NJL model we make use of the fact
that eqns (\ref{eharmosc}) and (\ref{eharmosc0})
imply the existence of operators $H^2$ and $H_0^2$ which have
the eigenvalues $\omega_i^2$ and $\omega_i^{(0)2}$ when
acting on the meson eigen--modes in the baryon number one and
zero sector, respectively. One may especially define the
``perturbation potential" $V$ via
\be
H^2=H_0^2+V.
\label{defpot}
\ee
A finite expression for the vacuum energy is obtained after
subtracting the first three terms in the expansion
\be
{\it Tr}\left(\sqrt{H_0^2+V}\right)={\it Tr}\left(H_0\right)
+\frac{1}{2}{\it Tr}\left(H_0^{-1}V\right)
-\frac{1}{8}{\it Tr}\left(H_0^{-1}VH_0^{-2}V\right)+\ldots
\label{exph2}
\ee
resulting in the finite energy correction due to the quantum
fluctuations \cite{ho94}
\be
\triangle E=
\frac{1}{2}{\it Tr}\left(H-H_0-\frac{1}{2}H_0^{-1}V
+\frac{1}{8}H_0^{-3}V^2\right).
\label{deltae1}
\ee
This expression, which was derived by Holzwarth in the case of the
Skyrme model, represents the actual starting point of our investigations
in the context of the NJL model. The main complication in comparison
with the Skyrme model is that we do not have explicit access to the
operators $H$ and $H_0$. However, we are able to compute the associated
energy eigenvalues by solving the Bethe--Salpeter equation
(\ref{bseqn}). Furthermore we have a suitable definition
(\ref{defmatel}) for the overlap of the eigenstates of $H$ and $H_0$.
The energy correction (\ref{deltae1}) actually represents the $3+1$
dimensional generalization of the result obtained by Cahill,
Comtet and Glauber \cite{ca76} in a $1+1$ dimensional model. It should,
however, be noted that the correction (\ref{deltae1}) is not free
of ordering ambiguities.

As already mentioned we have access to the eigenvalues of $H$ and $H_0$
only. Then the correction (\ref{deltae1}) may be expressed as
\be
\triangle E&=&\frac{1}{2}
\sum_{i}\Bigg\{\omega_i
-\frac{1}{8}\sum_j\omega_j^{(0)}
\left|\langle\tilde{\mbox{\boldmath $\eta$}}
(\mbox{\boldmath $r$},\omega_i)
|\tilde{\mbox{\boldmath $\eta$}}^{(0)}
(\mbox{\boldmath $r$},\omega^{(0)}_j)\rangle\right|^2
\label{deltae2} \\ &&\hspace{4cm}\times
\left[3+6\left(\frac{\omega_i}{\omega_j^{(0)}}\right)^2
-\left(\frac{\omega_i}{\omega_j^{(0)}}\right)^4\right]\Bigg\}.
\nonumber
\ee
Here the overlap $\langle\tilde{\mbox{\boldmath $\eta$}}
(\mbox{\boldmath $r$},\omega_i)|\tilde{\mbox{\boldmath $\eta$}}^{(0)}
(\mbox{\boldmath $r$},\omega^{(0)}_j)\rangle$ shows up since the trace
has to be performed in a common Hilbert space to chop off both
frequency sums in the same way. At this point it
should also have become clear that the formal considerations in
section 2 on the orthonormalization of the solutions to the
Bethe--Salpeter equation are in fact unavoidable. Eqn (\ref{deltae2})
obviously yields $\triangle E=0$ in the absence of the soliton,
{\it i.e.} when $\omega_i=\omega_i^{(0)}$ and
$\tilde{\mbox{\boldmath $\eta$}}
(\mbox{\boldmath $r$},\omega_i)=
\tilde{\mbox{\boldmath $\eta$}}^{(0)}
(\mbox{\boldmath $r$},\omega_i^{(0)})$.
We furthermore observe from eqn (\ref{deltae2}) that $\triangle E$
is of the order $N_C^0$ in agreement with the assertions made in the
introduction.

{}From eqn (\ref{deltae2}) it is obvious that the zero modes,
{\it i.e.} states with $\omega_i=0$ may lead to a sizable reduction of
the total energy $E+\triangle E$ since for these states $\triangle E$
is negative. The fact that for scattering states
$\omega_i\approx\omega_i^{(0)}$ demonstrates that the main contributions
to the quantum corrections is indeed due to the existence of zero modes
in the soliton background. This can also be understood in the context
of the phase--shift $\delta(k)$ expression for the Casimir energy,
$E_{\rm cas}\sim(-1/2\pi) \int dk \delta(k)$ up to counterterms which
render this integral finite for $k\rightarrow\infty$ \cite{ra82}.
According to Levison's theorem an additional $\pi$ has to be included
for each bound state. The only bound states in the background of the
NJL soliton are the zero modes. Thus the channels in which these modes
appear may considerably contribute to $E_{\rm cas}$. In the proceeding
section we will therefore concentrate on the channels which contain the zero
modes in the NJL model.

\bigskip
\stepcounter{chapter}
\leftline{\large\it 4. Zero Mode Channels}
\bigskip

Zero modes, {\it i.e.} solutions to the Bethe--Salpeter equation
(\ref{bseqn}) with $\omega_i=0$,  arise whenever the stationary
background field (the ``vacuum" seen by the meson fluctuations) breaks
a symmetry of the underlying theory. In that sense the zero modes are
Goldstone bosons. In the case of the chiral soliton the hedgehog
field configuration (\ref{chsol}) violates the rotational
and translational invariance. We therefore expect zero modes to
be associated with infinitesimal spatial rotations and translations
of the soliton. Due to the grand spin symmetry the zero mode
corresponding to infinitesimal iso--rotations is equivalent to the
one of the spatial rotations. Although the model is invariant under
axial rotations (for massless pions) a corresponding zero mode
does not exist since the infinitesimal axial rotation does not
leave the vacuum configuration $(M=m)$ invariant.

In order to identify the zero modes we expand the parametrization
(\ref{parafl}) up to linear order in
$\mbox{\boldmath $\eta$}(\mbox{\boldmath $r$},t)$
\be
M=m\left\{\xi_0^2+i\xi_0
\mbox{\boldmath $\eta$}\cdot\mbox{\boldmath $\tau$}\xi_0
\right\}+\ldots
=M_0+im\xi_0
\mbox{\boldmath $\eta$}\cdot\mbox{\boldmath $\tau$}\xi_0
+\ldots\quad .
\label{expm}
\ee
For the extraction of the formal structure of the zero modes we have
to identify the linear term with $[{\cal G},M_0]$. Here ${\cal G}$
refers to the generator of the symmetry transformation. For the
spatial rotation this gives
\be
\mbox{\boldmath $\eta$}_R(\mbox{\boldmath $r$})
={\rm sin}\Theta(r)\hat{\mbox{\boldmath $r$}}\times
\mbox{\boldmath $\delta$}_R
\label{etar}
\ee
where $\mbox{\boldmath $\delta$}_R$ is a measure for the infinitesimal
rotation. Similarly the translation defines
\be
\mbox{\boldmath $\eta$}_T(\mbox{\boldmath $r$})
=\Theta^\prime(r)\hat{\mbox{\boldmath $r$}}
\hat{\mbox{\boldmath $r$}}\cdot
\mbox{\boldmath $\delta$}_T+
\frac{{\rm sin}\Theta(r)}{r}\left(
\mbox{\boldmath $\delta$}_T-
\hat{\mbox{\boldmath $r$}}
\hat{\mbox{\boldmath $r$}}\cdot
\mbox{\boldmath $\delta$}_T\right).
\label{etat}
\ee
It is straightforward to verify that both
$\mbox{\boldmath $\eta$}_R$ and $\mbox{\boldmath $\eta$}_T$
carry unit grand spin, {\it i.e.} these are dipoles in grand spin
space. In accordance to the Skyrme model notation \cite{wa84} we
will refer to the channel which contains the rotational zero mode
as magnetic dipole ($M1$) while the channel with the translational
zero mode is called electric dipole ($E1$).

{}From the consideration of the zero modes we have obtained the
quantum numbers of the fluctuations in the $M1$ and $E1$ channels.
This allows us to make {\it ans\"atze}, which separate the radial and
angular dependencies, for general fluctuations in these channels
\be
\mbox{\boldmath $\tau$}\cdot\tilde{\mbox{\boldmath $\eta$}}_{M1}
(\mbox{\boldmath $r$},\omega)=
\mbox{\boldmath $\tau$}\cdot\left(
\hat{\mbox{\boldmath $r$}}\times
\mbox{\boldmath $\zeta$}(r,\omega)\right)
=\frac{i}{2}\left[
\mbox{\boldmath $\tau$}\cdot\mbox{\boldmath $\zeta$}(r,\omega),
\mbox{\boldmath $\tau$}\cdot\hat{\mbox{\boldmath $r$}}\right]
\label{ansatzm1}
\ee
and
\be
\mbox{\boldmath $\tau$}\cdot\tilde{\mbox{\boldmath $\eta$}}_{E1}
(\mbox{\boldmath $r$},\omega)=
\mbox{\boldmath $\tau$}\cdot\mbox{\boldmath $\zeta$}_A(r,\omega)
+\mbox{\boldmath $\tau$}\cdot\hat{\mbox{\boldmath $r$}}
\mbox{\boldmath $\tau$}\cdot\mbox{\boldmath $\zeta$}_B(r,\omega)
\mbox{\boldmath $\tau$}\cdot\hat{\mbox{\boldmath $r$}}.
\label{ansatze1}
\ee
Altogether these {\it ans\"atze} introduce nine radial functions.
Although the parametrization (\ref{ansatze1}) is not intuitively
clear from (\ref{etat}) it is the most convenient one since the
action of the grand spin zero object $\gamma_5\mbox{\boldmath $\tau$}
\cdot\hat{\mbox{\boldmath $r$}}$ on the
quark wave--functions $\tilde\Psi_\nu(\mbox{\boldmath $r$})$ is
well-known \cite{ka84} and does neither change grand spin nor
parity quantum numbers. This knowledge is also the reason why we
expressed $\tilde{\mbox{\boldmath $\eta$}}_{M1}$ in terms of a
commutator. In order to compute the matrix elements of the
{\it ans\"atze} (\ref{ansatzm1}) and (\ref{ansatze1}) we thus
only require the matrix elements of $\mbox{\boldmath $\tau$}$
times a radial function.

In the above notation the zero modes are parametrized by
\be
\mbox{\boldmath $\zeta$}^{\rm z.m.}(r)
&=&{\rm sin}\Theta(r)\ \mbox{\boldmath $\delta_R$}
\label{zzmrot} \\
\mbox{\boldmath $\zeta$}_A^{\rm z.m.}(r)
&=&\frac{1}{2}\left(\Theta^\prime(r)+
\frac{{\rm sin}\Theta(r)}{r}\right)
\mbox{\boldmath $\delta_T$}\ , \quad
\mbox{\boldmath $\zeta$}_B^{\rm z.m.}(r)
=\frac{1}{2}\left(\Theta^\prime(r)-
\frac{{\rm sin}\Theta(r)}{r}\right)\mbox{\boldmath $\delta_T$}\ .
\label{zzmtrans}
\ee

Now the main task is to substitute the {\it ans\"atze}
(\ref{ansatzm1}) and (\ref{ansatze1}) into the action functional.
Although these calculations are straightforward they are quite
tedious and we do not go into the details here. As a matter of
fact the presentation of the associated formulae would approximately
double the length of this paper. Let us rather display the generic form
of the Bethe--Salpeter equations for radial functions defined above.
As a matter of isospin invariance the action only depends on the
combinations $\mbox{\boldmath $\zeta$}(r,\omega)\cdot
\mbox{\boldmath $\zeta$}(r^\prime,-\omega)$ for the $M1$ channel
on the one hand and $\mbox{\boldmath $\zeta$}_A(r,\omega)\cdot
\mbox{\boldmath $\zeta$}_A(r^\prime,-\omega)$,
$\mbox{\boldmath $\zeta$}_A(r,\omega)\cdot
\mbox{\boldmath $\zeta$}_B(r^\prime,-\omega)$ as well as
$\mbox{\boldmath $\zeta$}_B(r,\omega)\cdot
\mbox{\boldmath $\zeta$}_B(r^\prime,-\omega)$ for the $E1$ channel on
the other hand. Thus the isospin invariance reduces the number of
independent radial functions from nine to three. We label these by
$\zeta(r,\omega)$, $\zeta_A(r,\omega)$ and $\zeta_B(r,\omega)$ and
keep in mind that they are three--fold degenerate.
This causes an overall factor $3$ in eqn (\ref{deltae2})
when one considers the $M1$ and $E1$ channels.
The Bethe--Salpeter equation for $\zeta(r,\omega)$ becomes
an homogeneous integral equation in the radial coordinate
\be
r^2\left\{
\int dr^\prime r^{\prime2}
\Phi_2^{M1}(r,r^\prime,\omega^2)\zeta(r^\prime,\omega)
+\Phi_1^{M1}(r)\zeta(r,\omega)\right\}=0
\label{bszm1}
\ee
while the Bethe--Salpeter equation for radial function in the
electric channel couples $\zeta_A(r,\omega)$ and $\zeta_B(r,\omega)$
\be
&&r^2\Bigg\{
\int dr^\prime r^{\prime2}
\left[\Phi_2^{E1AA}(r,r^\prime,\omega^2)\zeta_A(r^\prime,\omega)
+\Phi_2^{E1AB}(r,r^\prime,\omega^2)\zeta_B(r^\prime,\omega)\right]
\nonumber \\ && \hspace{4cm}
+\Phi_1^{E1}(r)\left[\zeta_A(r,\omega)-
\frac{1}{3}\zeta_B(r,\omega)\right]\Bigg\}=0
\label{bszea} \\
&&r^2\Bigg\{
\int dr^\prime r^{\prime2}
\left[\Phi_2^{E1BB}(r,r^\prime,\omega^2)\zeta_B(r^\prime,\omega)
+\Phi_2^{E1BA}(r,r^\prime,\omega^2)\zeta_A(r^\prime,\omega)\right]
\nonumber \\ && \hspace{4cm}
+\Phi_1^{E1}(r)\left[\zeta_B(r,\omega)-
\frac{1}{3}\zeta_A(r,\omega)\right]\Bigg\}=0
\label{bszeb}
\ee
Upon explicit computation it can be shown that the non--diagonal
elements of the Bethe--Salpeter kernel satisfy
$\Phi_2^{E1AB}(r,r^\prime,\omega^2)=
\Phi_2^{E1BA}(r^\prime,r,\omega^2)$ which, of course, reflects
the Hermitian character of the Bethe--Salpeter kernel. As a further
consequence the diagonal elements $\Phi_2^{M1}$ as well as
$\Phi_2^{E1AA}$ and $\Phi_2^{E1BB}$ turn out to be symmetric.

The numerical treatment of equations like (\ref{bszm1}) is described
at length in ref.\cite{we94a}. Thus we will only explain the key steps.
As the diagonalization (\ref{diagh}) of the Dirac Hamiltonian (\ref{h0})
is performed utilizing a spherical box of radius $D$ in order to
discretize the momentum eigenstates this geometry transfers to the
Bethe--Salpeter equation for the meson fluctuations. The radial
coordinate is then discretized ($r_k=\triangle r(k-1),\ k=1,...,N$
and $r_N=D$ determines $\triangle r$) which transforms
the integral equations into matrix equations. Then these matrix
equations are extended to eigenvalue equations \cite{we93} by setting
the $RHS$ to $\lambda(\omega)\zeta(\mbox{\boldmath $r$},\omega)$. The
eigenvalue $\lambda(\omega)$ then depends
on the frequency $\omega$. Finally $\omega$ is tuned to the
eigen--frequency $\omega_i$ such that $\lambda(\omega_i)=0$.
The associated eigenvector represents the eigen--wave--function in
the descretized form. The bilocal parts of the kernels only involve
matrix elements of $h_1$ (\ref{h12}). In these matrix elements
upper and lower components of the Dirac spinors are coupled because
$h_1$ is linear in $\gamma_5$. At $r=D$ the unitary transformation
${\cal T}$ (\ref{deft}) equals unity. Thus the integrands of the
matrix elements of $h_1$ at $r=D$ are linear combinations of terms
which are products of an upper and a lower component of the
eigen--functions of $h_0$ at the boundary,
$\Psi_\nu(|\mbox{\boldmath $r$}|=D)$. For reasons which will be
explained below we impose boundary conditions on these eigen--spinors
such that the upper components always vanish at $r=D$ \cite{we92}. It
is then obvious that the bilocal parts of the kernels are zero as one
of the arguments ($r$ or $r^\prime$) lies on the boundary. On the other
hand the local kernels have finite components at $r=D$. Thus the
boundary conditions for the Dirac spinors imply boundary conditions
for the meson fluctuations as well
\be
\zeta(D,\omega)=\zeta_A(D,\omega)=\zeta_B(D,\omega)=0.
\label{boundzeta}
\ee
For finite $D$ these conditions lead to discretized mesonic modes.

Since the algebraic expressions for the Bethe--Salpeter equations are
by far too complicated to recognize the appearance of the zero modes
we have to verify their existence numerically. In order to do so we
diagonalize the discretized kernels for $\omega_i=0$. Then we compare
the wave--functions associated with the lowest eigenvalues with the
radial functions given in eqns (\ref{zzmrot}) and (\ref{zzmtrans}).
The results are shown in figure 4.1 for the constituent
quark mass $m=400$MeV. The size of the spherical cavity is chosen to
be $D=6$fm, {\it i.e.} large compared to the typical extension of the
soliton (1fm).
For the rotational zero mode we find excellent agreement, while for
the translational zero mode a small deviation can be observed in
the vicinity of the origin. This, however, is due to the fact that
the numerical computation of the derivative $\Theta^\prime$ near
$r=0$ is burdened with some small errors. We should also mention that
the lowest eigenvalue of the Bethe--Salpeter kernel at $\omega=0$
is about three orders of magnitude smaller than the next to
lowest one. {\it I.e.} the radial functions displayed in figure
4.1 are in fact solutions to the equations
(\ref{bszm1}), (\ref{bszea}) and (\ref{bszeb}). Thus we have also
numerically established the existence of zero modes in the background
field of the NJL soliton. Furthermore this verification provides
an excellent check on our algebraical and numerical computations of the
kernels which are quite involved.

Considering figure 4.1 the sizable slope of
$\zeta_A^{\rm z.m.}(r)$ indicates that the translational zero mode has
non--negligible overlaps
$\langle\tilde{\mbox{\boldmath $\eta$}}_{E1}^{\rm z.m.}
(\mbox{\boldmath $r$})|\tilde{\mbox{\boldmath $\eta$}}^{(0)}
(\mbox{\boldmath $r$},\omega^{(0)}_j)\rangle$ with states of large
$\omega^{(0)}_j$. Thus one might run into problems concerning
the natural cut--off $\omega_{\rm thres}^{(0)}=2m$. We will
concentrate on this when discussing the numerical results for the
energy correction in the following section. First we have to
define the metric for the $E1$ and $M1$ channels. In order to
take account of the spherical symmetry we define the metric for
the $M1$ channel in the discretized coordinate space ($r_i$) via
\be
{\cal M}^{M1}_{kl}\omega_i)=
(\triangle r)^2\ r_k^2 r_l^{\prime2}\frac{\partial}{\partial\omega^2}
\Phi_2^{M1}(r_k,r_l^\prime,\omega)\Big|_{\omega=\omega_i}
\label{mm1}
\ee
while in the $E1$ channel
\be
{\cal M}^{E1}_{kl}(\omega_i)=(\triangle r)^2\ r_k^2 r_l^{\prime2}
\frac{\partial}{\partial\omega^2}
\pmatrix{\Phi_2^{E1AA}(r_k,r_l^\prime,\omega) &
\Phi_2^{E1AB}(r_k,r_l^\prime,\omega) \cr
\Phi_2^{E1BA}(r_k,r_l^\prime,\omega) &
\Phi_2^{E1BB}(r_k,r_l^\prime,\omega) \cr}\Bigg|_{\omega=\omega_i}
\label{me1}
\ee
acts in the $2N$--component vector space spanned by
$\pmatrix{\zeta_{Ak}(\omega_i)\cr\zeta_{Bk}(\omega_i)\cr}$. In the
discretized coordinate space these metric tensors are symmetric
matrices ${\cal M}^{M1,E1}_{kl}(\omega_i)$. In particular there exist
transformations $V$'s such that the $V^{\dag}{\cal M}V$'s are
diagonal. Then the roots (\ref{defsqrtm}) are defined via the
eigenvalues\footnote{These eigenvalues should not be mixed up with
the auxiliary eigenvalues, which were introduced to solve the
Bethe--Salpeter equation.} $\lambda_m^{M,E}(\omega_i)$ of
${\cal M}^{M1,E1}$
\be
\left(\sqrt{{\cal M}}\right)_{kl}(\omega_i)=
\left(V\cdot{\rm diag}\left(\sqrt{\lambda_1(\omega_i)},..,
\sqrt{\lambda_N(\omega_i)}\right)\cdot V^{\dag}\right)_{kl}
\label{rootmat}
\ee
for the $M1$ and $E1$ channels separately. In the latter case the
matrices are $2N\times2N$ dimensional. According to (\ref{defphi2})
the metric is defined into the wave--functions which solve the
Bethe--Salpeter equation
\be
\phi^{M1}_k(\omega_i)&=&\triangle r \sum_l
\left(\sqrt{{\cal M}}\right)_{kl}(\omega_i)\zeta_{k}(\omega_i)
\label{phim1} \\
\pmatrix{\phi^{E1}_{Ak}(\omega_i)\cr\phi^{E1}_{Bk}(\omega_i)\cr}
&=&\triangle r \sum_l
\pmatrix{\left(\sqrt{{\cal M}}\right)_{kl}^{AA}(\omega_i) &
\left(\sqrt{{\cal M}}\right)_{kl}^{AB}(\omega_i) \cr
\left(\sqrt{{\cal M}}\right)_{kl}^{BA}(\omega_i) &
\left(\sqrt{{\cal M}}\right)_{kl}^{BB}(\omega_i) \cr}\cdot
\pmatrix{\zeta_{Al}(\omega_i)\cr\zeta_{Bl}(\omega_i)\cr}.
\label{phie1}
\ee
These modified wave--functions are then subject to the trivial overall
normalization
\be
1=\triangle r \sum_k\phi^{M1}_k(\omega_i)^2\quad {\rm and}\quad
1=\triangle r \sum_k\left(\phi^{E1}_{Ak}(\omega_i)^2+
\phi^{E1}_{Bk}(\omega_i)^2\right).
\label{normphi}
\ee
Furthermore the overlaps are given by
\be
\hspace{-1cm}
\langle\tilde{\mbox{\boldmath $\eta$}}_{E1}
(\mbox{\boldmath $r$},\omega_i)|
\tilde{\mbox{\boldmath $\eta$}}^{(0)}
(\mbox{\boldmath $r$},\omega^{(0)}_j)\rangle&=&
\triangle r \sum_k
\phi^{M1}_k(\omega_i)\phi^{(0)M1}_k(\omega^{(0)}_j)
\label{overm1} \\
\hspace{-1cm}
\langle\tilde{\mbox{\boldmath $\eta$}}_{E1}
(\mbox{\boldmath $r$},\omega_i)|
\tilde{\mbox{\boldmath $\eta$}}^{(0)}
(\mbox{\boldmath $r$},\omega^{(0)}_j)\rangle&=&
\triangle r
\sum_k\left(\phi^{E1}_{Ak}(\omega_i)\phi^{(0)E1}_{Ak}(\omega^{(0)}_j)
+\phi^{E1}_{Bk}(\omega_i)\phi^{(0)E1}_{Bk}(\omega^{(0)}_j)\right).
\label{overe1}
\ee
Here $\phi^{(0)}$ represent the analogues of $\phi$ in the absence of
the soliton.

Although we are now completely equipped to compute the energy correction
$\triangle E$ we postpone this to the next section and rather add some
comments on the normalization of the zero modes.
One can easily show that substituting the rotational zero mode
(\ref{etar}) into $h_1$ (\ref{h12}) corresponds to
\be
h_1\left(\mbox{\boldmath $\eta$}_R\right)=
\frac{i}{2}\left[\mbox{\boldmath $\tau$}\cdot
\mbox{\boldmath $\delta$}_R,h_0\right]
\label{h1rzm}
\ee
with $h_0$ being the static Dirac Hamiltonian (\ref{h0}). Since, by
definition, the zero mode has $\omega_i=0$ one obtains for its
normalization from eqn (\ref{norm1})
\be
1&=&\frac{N_C}{2}\eta^{\rm val}\sum_{\mu\ne{\rm val}}
\frac{\langle{\rm val}|\mbox{\boldmath $\tau$}\cdot
\mbox{\boldmath $\delta$}_R|\mu\rangle\langle\mu|
\mbox{\boldmath $\tau$}\cdot
\mbox{\boldmath $\delta$}_R|{\rm val}\rangle}
{\epsilon_\mu-\epsilon_{\rm val}}
\nonumber \\ &&
+\frac{N_C}{8}\sum_{\mu\nu}
\langle\nu|\mbox{\boldmath $\tau$}\cdot
\mbox{\boldmath $\delta$}_R|\mu\rangle\langle\mu|
\mbox{\boldmath $\tau$}\cdot
\mbox{\boldmath $\delta$}_R|\nu\rangle
\left(\epsilon_\mu-\epsilon_\nu\right)^2
\int_{1/\Lambda^2}^\infty ds \sqrt{\frac{s}{4\pi}}\int_0^1 dx
\nonumber \\ && \hspace{1cm} \times
\left\{1-sx(1-x)\left(\epsilon_\mu+\epsilon_\nu\right)^2\right\}
{\rm exp}\left[-s\left((1-x)\epsilon_\mu^2
+x\epsilon_\mu^2\right)\right].
\label{rotzmnorm1}
\ee
The Feynman parameter integral in this equation can be carried out
resulting in
\be
1=\delta_R^a\delta_R^b\alpha^2_{ab}
\label{rotzmnorm2}
\ee
where
\be
\alpha^2_{ab}&=&\frac{N_C}{2}\eta^{\rm val}\sum_{\mu\ne{\rm val}}
\frac{\langle{\rm val}|\tau^a|\mu\rangle
\langle\mu|\tau^b|{\rm val}\rangle}
{\epsilon_\mu-\epsilon_{\rm val}}
\label{rotzmnorm3} \\ &&
+\frac{N_C}{4}\sum_{\mu\nu}\langle\nu|\tau^a|\mu\rangle
\langle\mu|\tau^b|\nu\rangle
\int_{1/\Lambda^2}^\infty\frac{ds}{\sqrt{4\pi s^3}}
%\nonumber \\ && \hspace{1cm} \times
\left\{\frac{e^{-s\epsilon_\nu^2}-e^{-s\epsilon_\mu^2}}
{\epsilon_\mu^2-\epsilon_\nu^2}
-s\frac{\epsilon_\nu e^{-s\epsilon_\nu^2}
+\epsilon_\mu e^{-s\epsilon_\mu^2}}
{\epsilon_\nu+\epsilon_\mu}\right\}
\nonumber
\ee
is just the moment of inertia for the chiral soliton \cite{re89} which
actually turns out to be an isoscalar,
$\alpha^2_{ab}=\alpha^2\delta^{ab}$.
Thus we have shown that the rotational zero mode is normalized with
respect to the moment of inertia. Analogously one can show that
the translational zero mode is normalized with respect to the
``pushing mass" tensor $E_{\rm push}\delta^{ab}$. This tensor is
defined by replacing the isospin generators, $\tau^a$ in eqn
(\ref{rotzmnorm3}) by the generators for the infinitesimal
translation, $i\partial_a$. In ref.\cite{po91} it has been demonstrated
that the pushing mass is identical to the energy of the static
soliton (\ref{ecl}), {\it i.e.} $E_{\rm push}=E_{\rm cl}$.
Thus the translational zero mode is indeed normalized with respect
to the classical mass of the soliton. We may reverse these results
and obtain a possibility to check our metric tensors
${\cal M}^{M1,E1}(\mbox{\boldmath $r$},
\mbox{\boldmath $r$}^\prime,\omega=0)$
\be
\int d^3r \int d^3r^\prime
\mbox{\boldmath $\zeta$}_R^{\rm z.m}(\mbox{\boldmath $r$})\cdot
{\cal M}^{M1}(\mbox{\boldmath $r$},
\mbox{\boldmath $r$}^\prime,\omega=0)\cdot
\mbox{\boldmath $\zeta$}_R^{\rm z.m}(\mbox{\boldmath $r$}^\prime)&=&
\alpha^2\ \mbox{\boldmath $\delta$}_R^2
\label{zzmnorm} \\
\int d^3r \int d^3r^\prime
\mbox{\boldmath $\zeta$}_T^{\rm z.m}(\mbox{\boldmath $r$})\cdot
{\cal M}^{E1}(\mbox{\boldmath $r$},
\mbox{\boldmath $r$}^\prime,\omega=0)\cdot
\mbox{\boldmath $\zeta$}_T^{\rm z.m}(\mbox{\boldmath $r$}^\prime)&=&
E_{\rm cl}\ \mbox{\boldmath $\delta$}_T^2
\label{zzenorm}
\ee
where the matrix structure of ${\cal M}^{E1}$ is not explicitly
shown. Eqns (\ref{zzmnorm}) and (\ref{zzenorm}) provide a possibility
to check our calculations in the sense that we first compute the
metric tensors for $\omega=0$ and evaluate the integrals on
the $LHS$ by substituting eqns (\ref{zzmrot}) and (\ref{zzmtrans}).
These results are then compared with the direct evaluations
of $E_{\rm cl}$ (\ref{ecl}) and $\alpha^2$ (\ref{rotzmnorm3}).
Numerically we observe deviations as small of 0.1\% when box sizes of
the order $D=6$fm are used and the explicit form in terms of the chiral
angle is substituted in the integrals. When the solutions to the
Bethe--Salpeter equations at $\omega=0$ are employed to evaluate
the integrals the error is somewhat larger because these solutions do
not exactly match the explicit forms at very large radial distances.
In any event this error has to be considered small and thus provides
an excellent verification of the correctness of our algebraical as
well as numerical manipulations which are rather involved.

The appearance of the moment of inertia also determined the above
mentioned choice for the boundary conditions on the Dirac spinors
$\Psi_\nu$. For other boundary conditions, {\it e.g.} those suggested
by Kahana and Ripka \cite{ka84}, the moment of inertia is plagued by isospin
violations of the order $1/D$ \cite{we92}.

\bigskip
\stepcounter{chapter}
\leftline{\large\it 5. Numerical Results}
\bigskip

In the previous section we have already presented one of our
main numerical results: We have verified the existence of rotational
and translational zero modes. Before concentrating on the results
for the energy subtraction $\triangle E$ we wish to add a few remarks
on the solutions in the $M1$ and $E1$ channels when no soliton is
present, {\it i.e.} $\Theta(r)\equiv0$. These solutions are important
for the overlaps $\langle\mbox{\boldmath $\eta$}^{\rm z.m.}
|\mbox{\boldmath $\eta$}^{(0)}\rangle$. The $M1$ channel comes with
unit orbital angular momentum, $l=1$; {\it i.e.} a $P$--wave. Thus
the corresponding solutions to the Klein Gordon equation which
satisfy the boundary conditions (\ref{boundzeta}) read
\be
\zeta_{\rm free}(r,\omega_i^{(0)})\propto j_1(q_i^1r)
\label{pfree}
\ee
where the $q_i^l$ make the $l^{\rm th}$ spherical Bessel function vanish
at the boundary, $j_l(q_i^lD)=0$. Furthermore
$\omega_i^{(0)}=\sqrt{m^2_\pi+(q_i^1)^2}$. In the $E1$ channel the
situation is somewhat more involved since $S$--wave solutions
\be
\zeta_{A,\rm free}(r,\omega_i^{(0)})\propto j_0(q_i^0r)\
{\rm and}\ \zeta_{B,\rm free}(r,\omega_i^{(0)})=0\quad
{\rm with}\quad\omega_i^{(0)}=\sqrt{m^2_\pi+(q_i^0)^2}
\label{sfree}
\ee
as well as $D$--wave solutions
\be
\zeta_{A,\rm free}(r,\omega_i^{(0)})\propto j_2(q_i^2r)\ {\rm and}\
\zeta_{B,\rm free}(r,\omega_i^{(0)})\propto 3j_2(q_i^2r)\quad
{\rm with}\quad\omega_i^{(0)}=\sqrt{m^2_\pi+(q_i^2)^2}\quad
\label{dfree}
\ee
exist. We have then computed the solutions to the Bethe--Salpeter
equation (\ref{bseqn}) in the absence of the soliton and compared
these results with the above suggested solutions to the
Klein--Gordon equation. In figure 5.1 we display a typical
solution in the $M1$ channel. Here we have chosen $m_\pi=0$. From
eqn. (\ref{bseqn}) we obtain for this solution the eigen--frequency
$\omega^{(0)}=360.1$MeV which reasonably well compares with 363.5MeV
as indicated by the root $q_3^1D$ of the spherical Bessel function.
Except for a small vicinity of $r=D$ the radial behavior of our
solution to eqn (\ref{bseqn}) matches that of the associated spherical
Bessel function $j_1(q_3^1r)$.
In figure 5.2 the same comparison is
performed for the $S$-- and $D$--wave solutions in the $E1$ channel.
Again the eigen--frequencies which are suggested by the solutions of
the Klein--Gordon equation are reproduced at the order of 1\% and the
radial dependencies of the solutions to eqn (\ref{bseqn}) reasonably well
agree with the corresponding Bessel functions. This is especially the
case for $r\le D/2$. Since we are interested in the overlap with the
zero modes, which are well localized ({\it cf}. figure 4.1)
the deviation from the Bessel functions at $r\approx D$ is negligible
because this region is not relevant for the overlap. Thus we may
approximate the solutions to eqn (\ref{bseqn}) in the absence of the
soliton by the spherical Bessel functions as suggested in eqns
(\ref{pfree}), (\ref{sfree}) and (\ref{dfree}). This represents a
major simplification because solving eqn (\ref{bseqn}) numerically is
very time consuming\footnote{The computation of the kernels
$\Phi_{1,2}^{E1}$ for a given frequency $\omega$ takes about 20h-cpu
on a HP9000/710 workstation. In this context it should be remarked
that the auxiliary eigenvalues of the Bethe--Salpeter kernel for
$\Theta\equiv0$ depend on $\omega$ rather strongly. It turns out that
$\omega$ has to be adjusted with an accuracy better than 0.1\% in order
to solve the Bethe--Salpeter equation.}. This is in particular the case
for the $E1$ channel since the solutions corresponding to the
roots $q_i^0$ and $q_i^2$ lie quite close and are difficult to
disentangle numerically.

We thus employ the following procedure to evaluate the matrix elements
$\langle\mbox{\boldmath $\eta$}^{\rm z.m.}
|\mbox{\boldmath $\eta$}^{(0)}\rangle$: We firstly reproduce the
zero mode wave--function from the Bethe--Salpeter equation in the
presence of the soliton and compute the associated metric tensor.
This then provides the modified wave--functions $\phi^{\rm z.m.}(r)$
in both the $M1$ and $E1$ channels according to eqns (\ref{phim1}) and
(\ref{phie1}). Next we compute the metric tensors in the absence of the
soliton for the frequencies $\omega_i^{(0)}$ determined by the roots of
the Bessel functions. Substituting the approximations (\ref{pfree}),
(\ref{sfree}) and (\ref{dfree}) into eqns (\ref{phim1}) and
(\ref{phie1}) in turn leads to the modified wave--functions
$\phi^{(0)}(r)$. After normalizing these wave--functions according
to eqn (\ref{normphi}) we are finally enabled to compute the relevant
matrix elements as prescribed in eqns (\ref{overm1}) and
(\ref{overe1}). In the course of these calculations we encounter one
further problem. Due to numerical errors some eigenvalues of the
metric tensors turn out to be negative yielding eqn (\ref{rootmat})
ill--defined. As a matter of fact the absolute values of these negative
eigenvalues turn out to be about three orders of magnitude smaller than
the relevant positive ones. In any event we do not expect a numerical
accuracy better than about 1\%. We therefore ignore these negative
eigenvalues. This in some sense defines a truncated norm. The validity
of this treatment can be judged by first normalizing
$\mbox{\boldmath $\eta$}$ subject to
$\int d^3r \int d^3r^\prime \mbox{\boldmath $\eta$}
(\mbox{\boldmath $r$})\cdot
{\cal M}(\mbox{\boldmath $r$},\mbox{\boldmath $r$}^\prime)\cdot
\mbox{\boldmath $\eta$}(\mbox{\boldmath $r$}^\prime)=1$.
Then the modified wave--functions $\mbox{\boldmath $\phi$}$ are
computed with the truncated norm. Numerically we then find the
normalization of $\phi$ to deviate from unity by less than 0.01\%.
This, of course, justifies the above truncation of the norm which
turned out to be necessary as a consequence of the numerical inaccuracy.

We have now completed the presentation of the methods and treatments
which serve as input to compute the energy correction $\triangle E$.
Needless to mention that we also substitute the roots of the Bessel
functions for $\omega_j^{(0)}$ in eqn (\ref{deltae2}). We have already
mentioned that, a as consequence of the non--confining NJL model, real
solutions only exist for $\omega_j^{(0)}\le2m$. Due to the treatment
in a finite box of radius $D$ the lowest quark energy is
$\sqrt{m^2+(\pi/D)^2}$ which increases the threshold energy
to $\omega^{\rm th}=2\sqrt{m^2+(\pi/D)^2}$ in the $M1$ channel
while in the $E1$ channel\footnote{Note that in the $E1$ channel $h_1$
has vanishing matrix elements in the grand spin zero subsystem of
the quark modes.}
$\omega^{\rm th}=\sqrt{m^2+(\pi/D)^2}+\sqrt{m^2+(q_1^1)^2}$.
We therefore truncate the sum (\ref{deltae2}) accordingly. In order to
judge this truncation we define the sum of overlaps for the zero modes
\be
{\cal S}=\sum_{\omega_j^{(0)}\le\omega^{\rm th}}
\left|\langle\tilde{\mbox{\boldmath $\eta$}}^{\rm z.m.}
(\mbox{\boldmath $r$})|\tilde{\mbox{\boldmath $\eta$}}^{(0)}
(\mbox{\boldmath $r$},\omega^{(0)}_j)\rangle\right|^2
\label{deftrunc}
\ee
which should approach unity if the model were insensible to the
truncation. Note that the number of meson modes which lie
below $\omega^{\rm th}$ decreases as $m_\pi$ increases.

In table \ref{ta_m1} the results for the rotational zero mode are
presented. We see that the sum of overlaps ${\cal S}$ is about 0.9
for all sets of parameter used. This number, of course, increases
with the constituent quark mass $m$ because the threshold grows
proportionally. Furthermore ${\cal S}$ is sufficiently close to
unity in order to conclude that the energy subtraction
$\triangle E\approx-(250-300)$MeV is reliable. Actually this is
about 100MeV smaller than the corresponding value in the Skyrme
model \cite{ho94}.
\begin{table}
\caption{\label{ta_m1}The quantum corrections to the soliton mass
due to the rotational zero mode. The size of the spherical
cavity is $D=6$fm.}
{}~\newline
\centerline{\tenrm\smalllineskip
\begin{tabular}{|c|c c c|c c c|}
\hline
& \multicolumn{3}{|c|}{$m_\pi=0$} &
\multicolumn{3}{c|}{$m_\pi=135$MeV} \\
$m$(MeV) & 400 & 500 & 600 & 400 & 500 & 600 \\
\hline
${\cal S}$
& 0.88 & 0.94 & 0.95
& 0.83 & 0.89 & 0.91 \\
$\triangle E$(MeV)
& -201 & -274 & -290
& -244 & -297 & -323 \\
\hline
\end{tabular}}
\end{table}
We also observe that the mass correction due to the rotational zero
mode is about (30-40)MeV lower in the chiral limit than for the physical
value of the pion mass, $m_\pi=135$MeV.

Table \ref{ta_e1} contains the results for the translational zero mode.
We see that the sum of overlaps is significantly smaller than for the
rotational zero mode (table \ref{ta_m1}). Here it hardly reaches 0.5.
This is due to the fact that the slope of the profile function
corresponding to the translational zero mode is enhanced compared to
that of the rotational zero mode. Thus a sizable number of Fourier
components with non--vanishing overlap lie beyond the
quark--antiquark threshold.
\begin{table}
\caption{\label{ta_e1}The quantum corrections to the soliton mass due
to the translational zero mode. The contributions stemming from the
$S(l=0)$-- and $D(l=2)$--waves are disentangled. The size of the
spherical cavity is $D=6$fm.}
{}~\newline
\centerline{\tenrm\smalllineskip
\begin{tabular}{|c|c c c|c c c|}
\hline
& \multicolumn{3}{|c|}{$m_\pi=0$} &
\multicolumn{3}{c|}{$m_\pi=135$MeV} \\
$m$(MeV) & 400 & 500 & 600 & 400 & 500 & 600 \\
\hline
${\cal S}$
& 0.41 & 0.41 & 0.61
& 0.28 & 0.37 & 0.46 \\
$\triangle E_{l=0}$(MeV)
& -18  & -22  & -28
& -12  & -22  & -28 \\
$\triangle E_{l=2}$(MeV)
& -127 & -140 & -207
& -82  & -128 & -187 \\
$\triangle E$(MeV)
& -145 & -162 & -235
& -94  & -150 & -215 \\
\hline
\end{tabular}}
\end{table}
Hence we have to interpret the resulting $\triangle
E\approx-(100-200)$MeV as a lower bound for the energy subtraction
originating from the translational zero mode. Nevertheless we can
extract some qualitative statements from the results listed in
table \ref{ta_e1}. First of all we observe that the $D$--wave
contributions are dominating. This result is also found in the Skyrme
model \cite{ho94}. Furthermore we see that the value of the pion mass
has only little influence on the $S$--wave contribution while the
absolute value of the $D$--wave contribution decreases with increasing
pion mass. This is in contrast to the rotational zero mode. Thus the
net effect of varying the pion mass is somewhat mitigated.

Next we wish to estimate the contributions of the scattering states,
$\omega_i\ge m_\pi$, to the energy correction $\triangle E$ as
described by eqn (\ref{deltae2}). In order to do so we first have to
construct the corresponding solutions to the Bethe--Salpeter equation
(\ref{bseqn}). In table \ref{ta_scat} the frequencies of the lowest
solutions are compared to their analogues in the absence of the
soliton. In the $E1$ channel we distinguish between solutions
which are dominantly $S$-- or $D$--waves. We establish that the
scattering states always lie slightly above the associated states
for $\Theta=0$.
\begin{table}
\caption{\label{ta_scat}The contribution of the first scattering
states to the quantum corrections of the soliton mass in the
channel of the rotational ($M1$) and translational ($E1$) zero modes.
The scattering state under consideration is labeled $\eta$.
The entry ``max(..)" represents the contribution to ${\cal S}$
by the dominant term. The constituent quark and the pion masses
are $600$MeV and $135$MeV, respectively. The size of the spherical
cavity is $D=6$fm.}
{}~\newline
\centerline{\tenrm\smalllineskip
\begin{tabular}{|c|c|c|c|}
\hline
& $M1$ & \multicolumn{2}{|c|}{$E1$} \\
\cline{2-4}
& $l=1$  & $l=0$ & $l=2$ \\
\hline
$\omega^{(0)}$(MeV) & 202 & 171 & 235 \\
$\omega$(MeV) & 210 & 176 & 240 \\
max $\left(\langle\mbox{\boldmath $\eta$}^{(0)}_i|
\mbox{\boldmath $\eta$}\rangle^2\right)$
& 0.96 & 1.00 & 0.99 \\
${\cal S}$ & 0.99 & 1.00 & 1.00 \\
$\triangle E$(MeV)
& -1.9 &-0.6 & -0.5 \\
\hline
\end{tabular}}
\end{table}
We furthermore observe from table \ref{ta_scat} that the sum over
the overlaps ${\cal S}$ is strongly dominated by only one term. As a
matter of fact it is always the one which happens to have the same
number of knots in the radial part of the wave--function associated
with scattering state under consideration. This dominance causes
${\cal S}$ to very closely approach unity. Thus we conclude that
the contribution of the low--lying scattering states to $\triangle E$
of the order of only a few MeV is very reliable. We have found the same
result for the second scattering state in the $M1$ channel. For the
$E1$ channel the extraction of higher scattering states is somewhat
troublesome because the states which are dominantly $S$-- or $D$--waves
are almost degenerate. Nevertheless the results found so far for the
scattering states suggest that their contribution to $\triangle E$ is
almost negligible. The fact that these contributions turn out to be
negative is a consequence of the expansion (\ref{exph2}) when the
``perturbation" $V$ is attractive \cite{ho94a}. These results
qualitatively agrees with those obtained in the Skyrme model \cite{ho94}.

We are finally enabled to present our predictions for the masses
of baryons in the NJL soliton approach. For the projection onto
good quantum numbers of spin and isospin we employ the well--established
semi--classical cranking approach \cite{ad83}. This yields the mass
formula for a baryon of spin $J$
\be
M=E_{\rm cl}+\triangle E+\frac{J(J+1)}{2\alpha^2}.
\label{mbaryon}
\ee
According to the above discussions $\triangle E$ is understood
as the sum of the zero mode contributions listed in tables
\ref{ta_m1} and \ref{ta_e1}. In eqn (\ref{mbaryon}) $\alpha^2$ refers
to the moment of inertia (\ref{rotzmnorm3},\ref{zzmnorm}) \cite{re89}.
The individual pieces in eqn (\ref{mbaryon}) are of the
orders ${\cal O}(N_C)$, ${\cal O}(1)$ and ${\cal O}(1/N_C)$,
respectively. Here we have ignored the quantum corrections
at order ${\cal O}(1/N_C)$. These have been estimated in the
Skyrme model to be less than 100MeV \cite{ho94}. Hence their
contribution seems to be less than the uncertainties in our
estimates for $\triangle E$.
\begin{table}
\caption{\label{ta_baryon}The predictions for the masses of the
nucleon ($N$) and $\Delta$--resonance. The empirical data are
$939$MeV and $1232$MeV, respectively.}
{}~\newline
\centerline{\tenrm\smalllineskip
\begin{tabular}{|c|c c c|c c c|}
\hline
& \multicolumn{3}{|c|}{$m_\pi=0$} &
\multicolumn{3}{c|}{$m_\pi=135$MeV} \\
$m$(MeV) & 400 & 500 & 600 & 400 & 500 & 600 \\
\hline
$E_{\rm cl}$(MeV)  & 1212 & 1193 & 1166 & 1250 & 1221 & 1193\\
$\triangle E$(MeV) & -346 & -436 & -525 & -338 & -448 & -538\\
$\alpha^2$(1/GeV)  & 6.26 & 4.73 & 3.87 & 5.80 & 4.17 & 3.43\\
$M_N$(MeV)         & 926  & 836  & 738  & 976  & 863  & 764 \\
$M_\Delta$(MeV)    & 1166 & 1153 & 1126 & 1236 & 1223 & 1201\\
\hline
\end{tabular}}
\end{table}

We indeed find that the numerical results roughly follow this $1/N_C$
counting pattern. Unfortunately the results shown in table
\ref{ta_baryon} suggest that the overall prediction for the baryon
masses in the NJL soliton model underestimates the empirical value for
the nucleon mass ($939$MeV). Only for the constituent quark mass
$m\approx400$MeV a good description is obtained. In that case, however,
${\cal S}$ is as small as 0.4 in the $E1$ channel ({\it cf}.
table \ref{ta_e1}). Then, of course, the question of reliability
cannot be answered unambiguously. Although one may assume the point
of that this is a shortcoming of the NJL model in general rather
than only for the meson fluctuations. Hence one would have to
consider the results listed in tables \ref{ta_m1} and \ref{ta_e1}
seriously and conclude that the masses of the nucleon and
$\Delta$--resonance are reasonably well reproduced for
$m\approx400$MeV. As $m$ is further increased the
classical soliton energy $E_{\rm cl}$ as well as the quantum
corrections decrease; thereby underestimating the nucleon mass.
Simultaneously the $1/N_C$ counting pattern breaks down since also
the moment of inertia, $\alpha^2$, decreases leading to a
$\Delta$--nucleon mass splitting of the order of $\triangle E$. These
two quantities are, however, supposed to differ by one order in
$N_C$. In turn this renders the prediction for the mass of the
$\Delta$--resonance almost independent of the parameters and in
reasonable agreement with experimental value of $1232$MeV when the
physical value of the pion mass is adopted.

\bigskip
\stepcounter{chapter}
\leftline{\large\it 6. Conclusions}
\bigskip

In the present paper we have studied the quantum corrections,
$\triangle E$, to the classical mass, $E_{\rm cl}$, of the chiral
soliton in the NJL model. In a first step we have investigated the
formal structure of the Bethe--Salpeter equation for pionic
fluctuations in the soliton background. As the soliton is static
and the eigen--modes of this equation appear in pairs $\pm\omega_i$
we were able to demonstrate that the energy operator in the Fock
space of the meson fluctuations is identical to that of a harmonic
oscillator; although the time derivative operator appears at all
(even) orders in the coordinate space representation of the
Bethe--Salpeter equation. This result has allowed us to adopt the
expression for the quantum correction which was previously obtained
in the Skyrme model \cite{ho94}. According to our examinations this
expression is applicable to all static soliton models in $3+1$
dimensions as long as the eigen--modes occur in pairs. However, here
the overlap matrix element between solutions of the Bethe--Salpeter
equation with and without the soliton present, has turned out to be more
involved than in the Skyrme model in particular because the solutions
in the absence of the soliton do not exactly obey the Klein--Gordon
equation. Furthermore the sum over the free eigen--modes had to be
truncated since these modes are unstable against the decay into a
quark--antiquark pair once the corresponding eigen--frequency has
exceeded the threshold, $2m$.

Two types of eigen--modes exist in the presence of the soliton: The
bound zero modes and the scattering states. We have made plausible that
the latter type only contributes very little to $\triangle E$. Hence
the scattering modes have been discarded and $\triangle E$ has
completely been approximated by the zero mode contribution. This
procedure is also justified by the similar calculations in the Skyrme
model \cite{ho94}.

Before actually computing the quantum corrections we have, for the
first time, verified the existence of rotational and translation
zero modes in the background of the NJL soliton by explicit
construction. For the rotational zero mode the truncation caused
by the non--confining nature of the NJL model of the
mode sum does not represent a serious problem because the sum of
overlaps approximates unity fairly well. This in turn leads to a
reliable estimate for the quantum correction due to the rotational
zero mode of about $-250$MeV. In the case of the translational zero
mode the situation is somewhat worse because the overlaps sum up to
only about 0.5. In this sense one has to regard
$\triangle E\approx-150$MeV only as a bound.

One might, however, adopt a different point of view and consider the
above quoted data as the actual results. In any event the NJL model is
not well defined for frequencies above the quark--antiquark threshold.
All other quantities ({\it e.g.} the classical mass $E_{\rm cl}$) might
undergo significant changes as well once the model is improved to avoid
this problem. In this interpretation we have observed that the NJL model
predicts the masses of the nucleon and the $\Delta$ resonance fairly
well for a constituent quark mass $m\approx400$MeV. Even the $1/N_C$
counting scheme $E_{\rm cl}\sim N_C\triangle E\sim N_C^2(M_\Delta-M_N)$
seems to operate then. Upon increasing the constituent quark mass this
is no longer the case and the estimate for the nucleon mass turns out
to be somewhat too small while the prediction for $M_\Delta$ remains
almost unaltered. The smallness of the absolute value for the nucleon
mass appears to be connected to the fact that $E_{\rm cl}$ decreases
as the constituent quark mass  increases. This seems to be special
to the NJL model without the isoscalar--vector $\omega$--meson
included. Recently it has been demonstrated that the proper
incorporation of this field indeed yields a classical energy which
increases with $m$ \cite{we94b}. Thus one might suspect that the
associated extension of the NJL soliton model leads to an even more
reasonable description of the nucleon mass for larger constituent
quark masses. This would in a sense be more reliable because the sum
of overlap matrix elements would come closer to unity.

As a side--product we have been able to define a metric for the overlaps
of the meson states with and without the soliton present. As has
previously been pointed out in the context of the Skyrme model
\cite{ho90} this metric plays an important role for the computation of
the momentum dependent pion--nucleon form factor $g_{\pi NN}(q^2)$.
In order to compute $g_{\pi NN}(q^2)$ one na\"\i ely would
Fourier--transform ${\rm sin}\Theta(r)$. Commonly in soliton models
this procedure underestimates the cut-off
$\Lambda_{\pi NN}\approx1.6$GeV, which is defined via
\be
\frac{g_{\pi NN}(q^2<0)}{g_{\pi NN}(m_\pi^2)}=
\frac{\Lambda_{\pi NN}^2-m_\pi^2}{\Lambda_{\pi NN}^2-q^2}
\label{gpnn}
\ee
by a factor two or even more \cite{co86}. This may be improved by
the proper incorporation of the metric tensor because the
Fourier--transformation of ${\rm sin}\Theta(r)$ basically represents
the projection of the rotational zero mode onto free pion states.
Investigations in this direction in the context of the NJL model
are subject to future studies.

\vskip3cm

\leftline{\large\it Acknowledgement}
\bigskip

The authors gratefully acknowledge stimulating discussions with
G. Holzwarth on the quantum corrections in soliton models.

%\vfill\eject

\vfil\eject

\centerline{\Large \bf Figure captions}

\vskip1cm

\centerline{\large \bf Figure 4.1}

\noindent
Comparison of the numerical solution of the Bethe--Salpeter
equations (4.8,4.9,4.10) with the analytic form for the zero modes
as given in eqns (4.6) and (4.7). The lattice points are indicated.

\vskip1cm

\centerline{\large \bf Figure 5.1}

\noindent
Comparison of the numerical solution of the
Bethe--Salpeter equations (2.16) for $\Theta=0$ with
the corresponding spherical Bessel function in the $M1$ channel. Both
curves refer to the wave--function of the second excited state. The
normalization of the radial functions is arbitrarily chosen. The
lattice points are indicated.

\vskip1cm

\centerline{\large \bf Figure 5.2}

\noindent
Comparison of the numerical solution of the
Bethe--Salpeter equations (2.16) for $\Theta=0$ with
the corresponding spherical Bessel function in the $E1$ channel.
Left: The $S$--wave solution with $\zeta_B(r)\equiv0$. The
normalization of the radial functions is arbitrarily chosen. The
corresponding solution of the Klein--Gordon eqn has
$\omega^{(0)}=314.2$MeV. Right: The $D$--wave solution. Here
the linear combination $\zeta_A(r)-\zeta_B(r)/3$ vanishes. The
Klein--Gordon eqn yields the eigen--frequency $303.2$MeV.

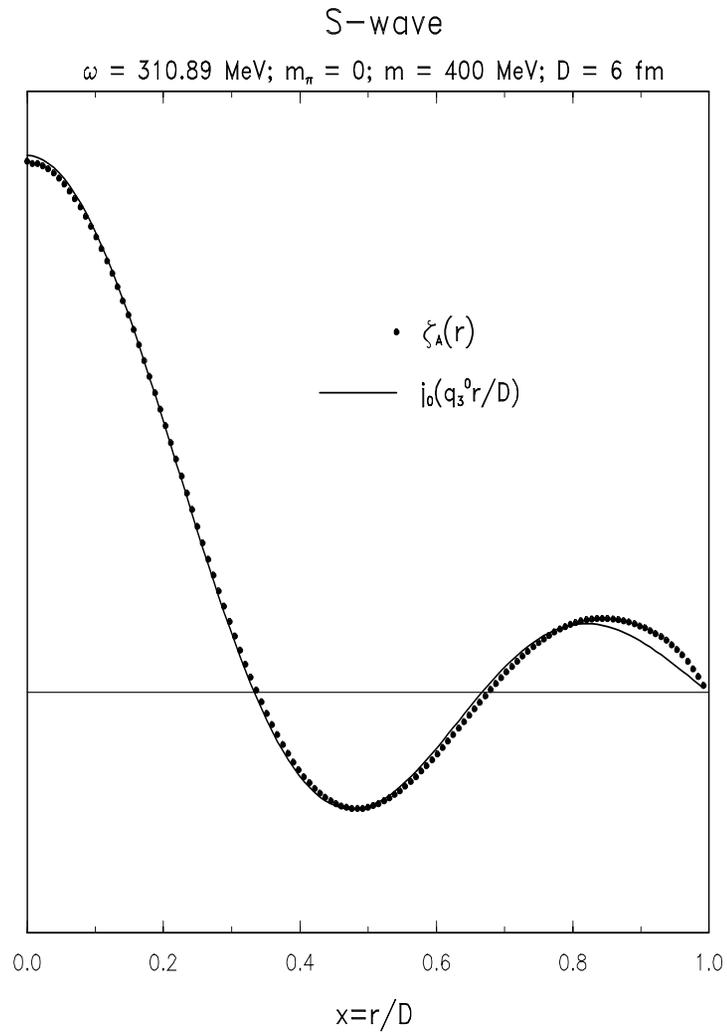
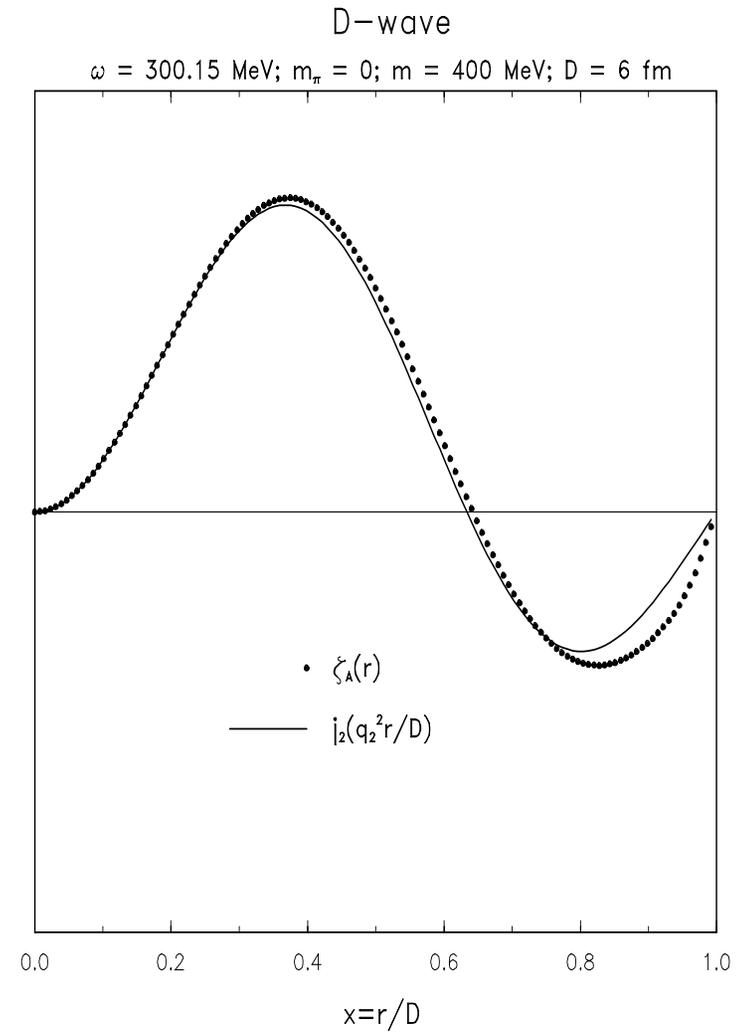
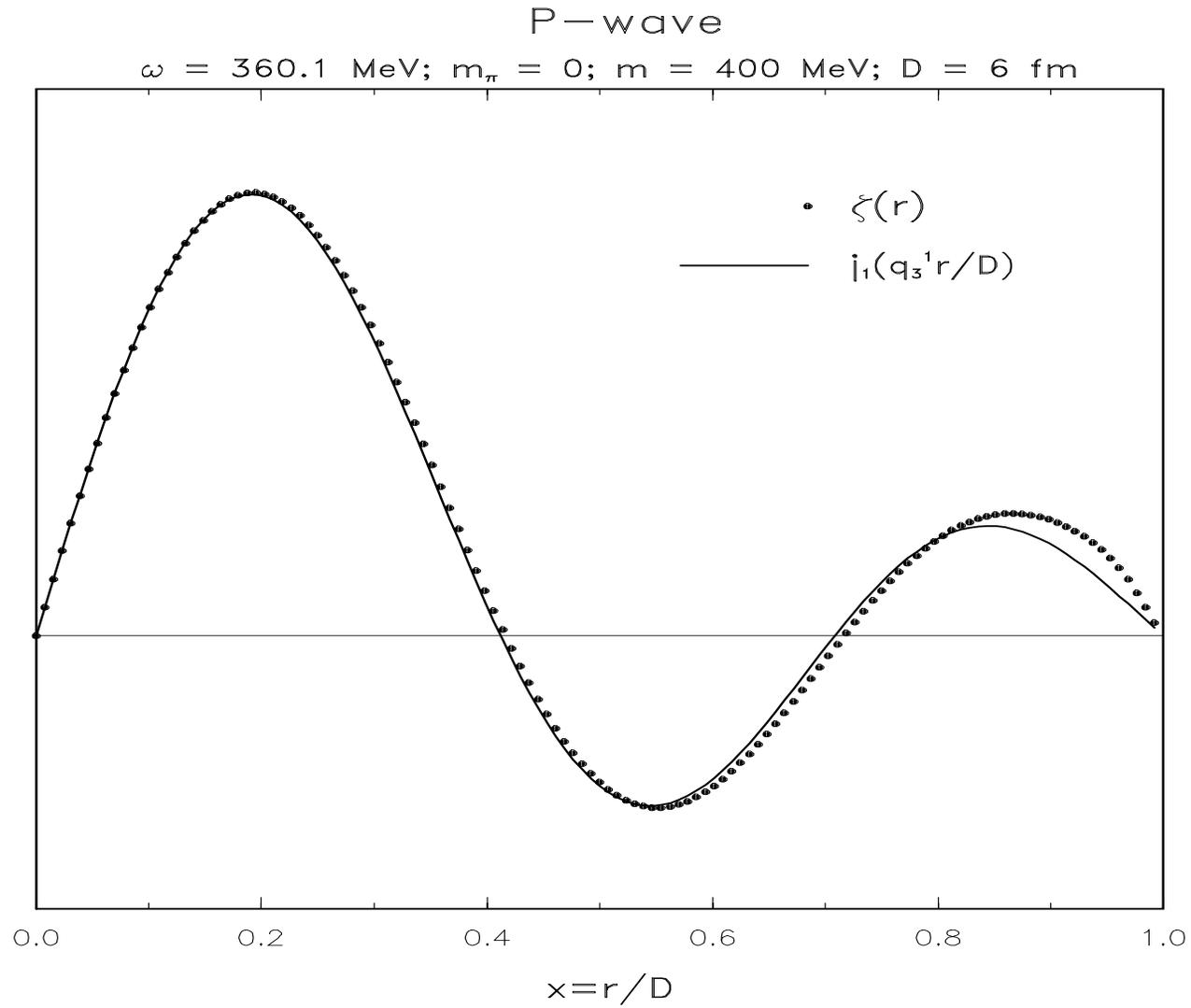
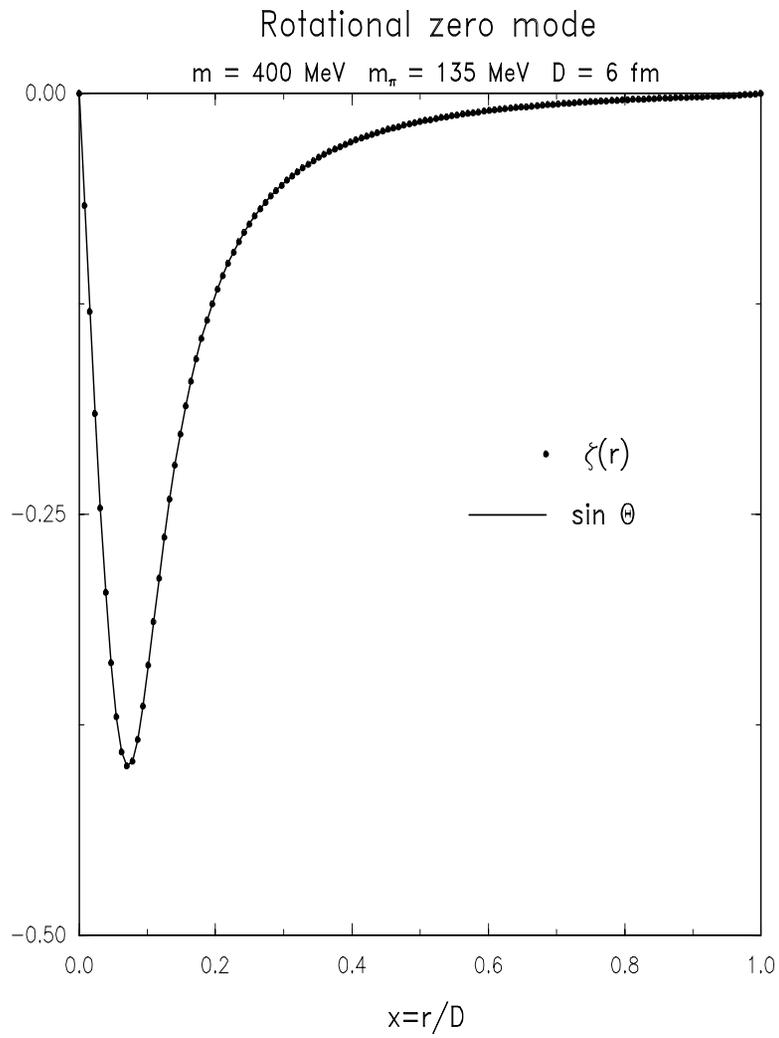
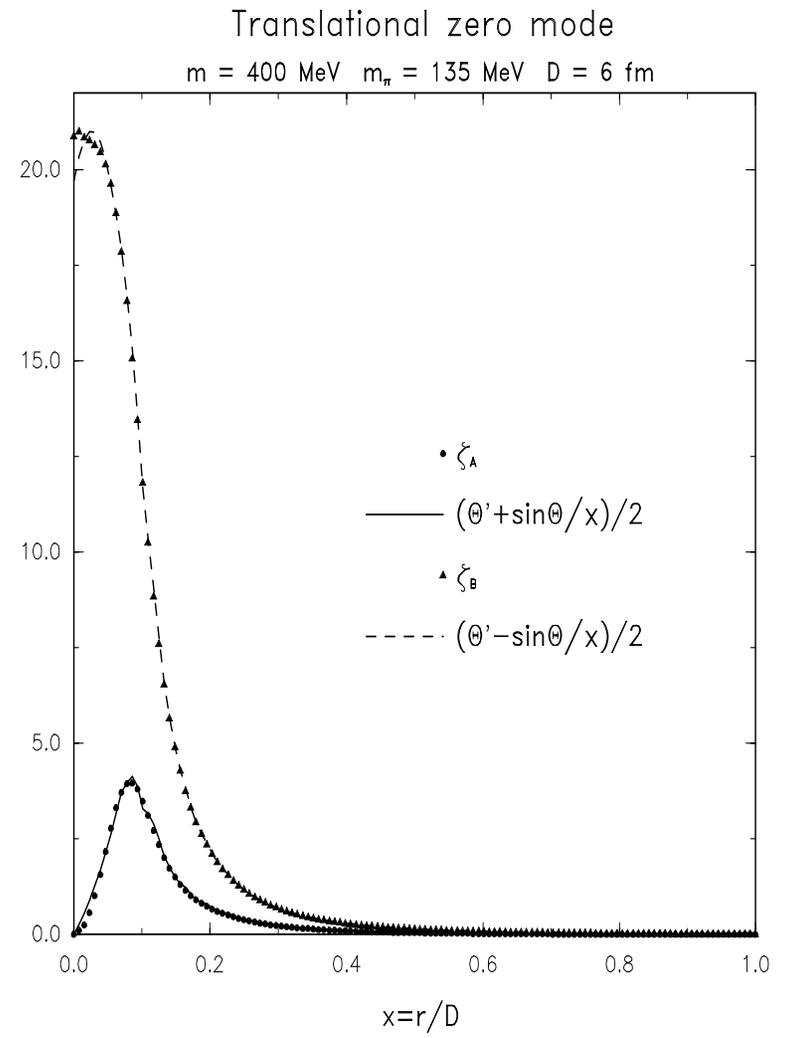
\end{document}